\documentclass{llncs}
\usepackage{amssymb}
\usepackage{amsmath}

\usepackage{color}
\usepackage{url}
\usepackage{ifpdf}
\usepackage{graphicx}
\usepackage{lipsum}
\usepackage{multirow}
\usepackage{booktabs}
\usepackage{times}
\usepackage{pgfplots} 
\usepackage{wrapfig}
\usepackage{xspace}
\usepackage[T1]{fontenc} 
\usepackage{multirow}
\ifpdf
\usepackage[colorlinks=true,linkcolor=blue,citecolor=blue]{hyperref}
\fi

\usepackage{verbatim}

\long\def\comment#1{\relax}
\newcommand{\eg}{{e.g.}}
\newcommand{\ie}{{i.e.}}
\newcommand{\etal}{\emph{et al.}}

\newcommand{\Sscr}{\mathcal{S}}

\newcommand{\tup}[1]{\langle#1\rangle}

\renewenvironment{paragraph}[1]{\smallskip\noindent{\it #1}~}

\usepackage{tweaklist}
\renewcommand{\enumhook}{
	\setlength{\topsep}{0em}
	\setlength{\itemsep}{0em}
	\setlength{\partopsep}{0em}
	\setlength{\parskip}{0em}
	\setlength{\parsep}{0em}
	\addtolength{\leftmargin}{-0.75em}
}
\renewcommand{\itemhook}{
	\setlength{\topsep}{0em}
	\setlength{\itemsep}{0em}
	\setlength{\partopsep}{0em}
	\setlength{\parskip}{0em}
	\setlength{\parsep}{0em}
	\addtolength{\leftmargin}{-0.75em}
}

\newcommand\lra{\longrightarrow}

\newcommand\RAP{RTA\xspace}
\newcommand\RAPs{RTAs\xspace}

\newcommand{\safer}{{\ensuremath{\mathsf{safer}}}}
\newcommand{\safe}{{\ensuremath{\mathsf{safe}}}}
\newcommand{\unsafe}{\ensuremath{{\mathsf{unsafe}}}}
\newcommand{\bad}{\ensuremath{{\mathsf{bad}}}}

\newcommand{\gapsafer}{\ensuremath{\mathsf{gap_{safer}}}}
\newcommand{\gapsafe}{\ensuremath{\mathsf{gap_{safe}}}}
\newcommand{\gapunsafe}{\ensuremath{\mathsf{gap_{unsafe}}}}
\newcommand{\dist}{\ensuremath{\mathsf{dist}}}

\newcommand{\vehl}{\ensuremath{\mathsf{veh_l}}}
\newcommand{\vehf}{\ensuremath{\mathsf{veh_f}}}

\newcommand{\pedA}{\ensuremath{\mathsf{st}}}
\newcommand{\pedB}{\ensuremath{\mathsf{fn}}}
\newcommand{\crossSt}{\ensuremath{\mathsf{cr_{1}}}}
\newcommand{\crossFn}{\ensuremath{\mathsf{cr_{2}}}}
\newcommand{\veh}{\ensuremath{\mathsf{veh}}}
\newcommand{\vel}{\ensuremath{\mathsf{v}}}
\newcommand{\acc}{\ensuremath{\mathsf{\alpha}}}
\newcommand{\pos}{\ensuremath{\mathsf{pos}}}
\newcommand{\px}{\ensuremath{\mathsf{px}}}
\newcommand{\py}{\ensuremath{\mathsf{py}}}
\newcommand{\pvel}{\ensuremath{\mathsf{v_p}}}

\newcommand{\maxDec}{\ensuremath{\mathsf{maxDec}}}

\newcommand{\dstop}{\ensuremath{\mathsf{dStop}}}

\newcommand\SA{\ensuremath{\mathsf{SA}}\xspace}
\newcommand\lkb{\ensuremath{\mathsf{lkb}}\xspace}
\newcommand{\SP}{{\ensuremath{\mathsf{SP}}}\xspace}
\newcommand{\LS}{{\ensuremath{\mathsf{LS}}}\xspace}
\newcommand{\conf}{{\ensuremath{\mathsf{conf}}}\xspace}

\newcommand{\react}{{\ensuremath{\mathsf{react}}}\xspace}
\newcommand{\maxacc}{{\ensuremath{\mathsf{maxacc}}}\xspace}
\newcommand{\maxdec}{{\ensuremath{\mathsf{maxdec}}}\xspace}
\newcommand{\dis}{{\ensuremath{\mathsf{dis}}}\xspace}

\newcommand\ErrSensor{{\ensuremath{\mathsf{ErrSensor}}}\xspace}
\newcommand\lowSpd{\ensuremath{\mathsf{lowSpd}}}
\newcommand\farAway{\ensuremath{\mathsf{farAway}}}
\newcommand{\drss}{{\ensuremath{\mathsf{drss}}\xspace}}

\long\def\omitthis#1{\relax}

\newcommand\dt{\ensuremath{\mathsf{dt}}\xspace}

\newcommand\tasks{\ensuremath{\mathsf{tasks}}\xspace}

\catcode`@=11 % borrow the private macros of PLAIN (with care)
\def\ldisplayindent{\quad}
\def\ldisplaylines#1{\displ@y\openup\jot
  \halign{\hbox to\displaywidth{$\ldisplayindent\displaystyle##\hfil$}\crcr
    #1\crcr}}

\def\ldisplaytwo#1{\displ@y\openup\jot
  \halign to\displaywidth{\rm ##\quad \tabskip\z@skip 
       &$\displaystyle{{}##}$\hfil\tabskip=0pt plus 1000pt minus 1000pt\crcr
    #1\crcr}}

\catcode`@=12 % at signs are no longer letters

\title{Technical-Report: Automating Recoverability Proofs for Cyber-Physical Systems with Runtime Assurance Architectures}

\author{Vivek Nigam\inst{2,3} and Carolyn Talcott\inst{1}}
\institute{
SRI International, Menlo Park, USA, \email{carolyn.talcott@gmail.com} 
\and
Federal University of Para\'iba, Jo\~ao Pessoa, Brazil,
\email{vivek.nigam@gmail.com}
\and 
Huawei Munich Research Center, Germany
}

\begin{document}
\maketitle
\pagestyle{plain} 
 % \linenumbers

\begin{abstract}
Cyber-physical systems (CPSes), such as autonomous vehicles, use sophisticated components like ML-based controllers. It is difficult to provide evidence about the safe functioning of such components.
To overcome this problem, Runtime Assurance Architecture (\RAP) solutions have been proposed. 
The \RAP's decision component evaluates the system's safety risk and whenever the risk is higher than acceptable the \RAP switches to a safety mode that, for example, activates a controller with strong evidence for its safe functioning.
% An \RAP pairs a verified safety component with the unverified one through a decision sub-component which is responsible for monitoring the CPS safety risk and switching to the verified controller when risk is high and to the unverified controller when risk is low.
In this way, \RAPs increase CPS runtime safety and resilience by recovering the system from higher to lower risk levels.
The goal of this paper is to automate recovery proofs of CPSes using \RAPs.
We first formalize the key verification problems, namely, the decision sampling-time adequacy problem and the time-bounded recoverability problem.
We then demonstrate how to automatically generate proofs for the proposed verification problems using symbolic rewriting modulo SMT.
Automation is enabled by integrating the rewriting logic tool (Maude), which generates sets of non-linear constraints, with an SMT-solver (Z3) to produce proofs
\end{abstract}

% \textbf{Keywords: runtime assurance, proof automation, symbolic rewriting, SMT solving, recoverability} 

\section{Introduction}
\label{sec:intro}
%!TEX root = tase23.tex

Cyber-physical systems (CPSes) are increasingly performing complex safety-critical missions in an autonomous fashion,  
autonomous vehicles (AVs) being a current prime example.
Given the complexity of the environment in which such CPSes operate, 
they often rely on highly complex machine learning (ML) based controllers~\cite{apollo} because of ML's capability of learning implicit requirements about the vehicle operation conditions. 
It has been notably hard, however, to provide safety arguments using only such ML-based components due to their functional insufficiency~\cite{sotif}. Despite the great amount of effort in building methods for verifying systems with ML-based components, they still present more faults than acceptable~\cite{rushby20safecomp}.

\begin{wrapfigure}{r}{0.55\textwidth}
  \vspace{-5mm}
    \includegraphics[width=0.55\textwidth]{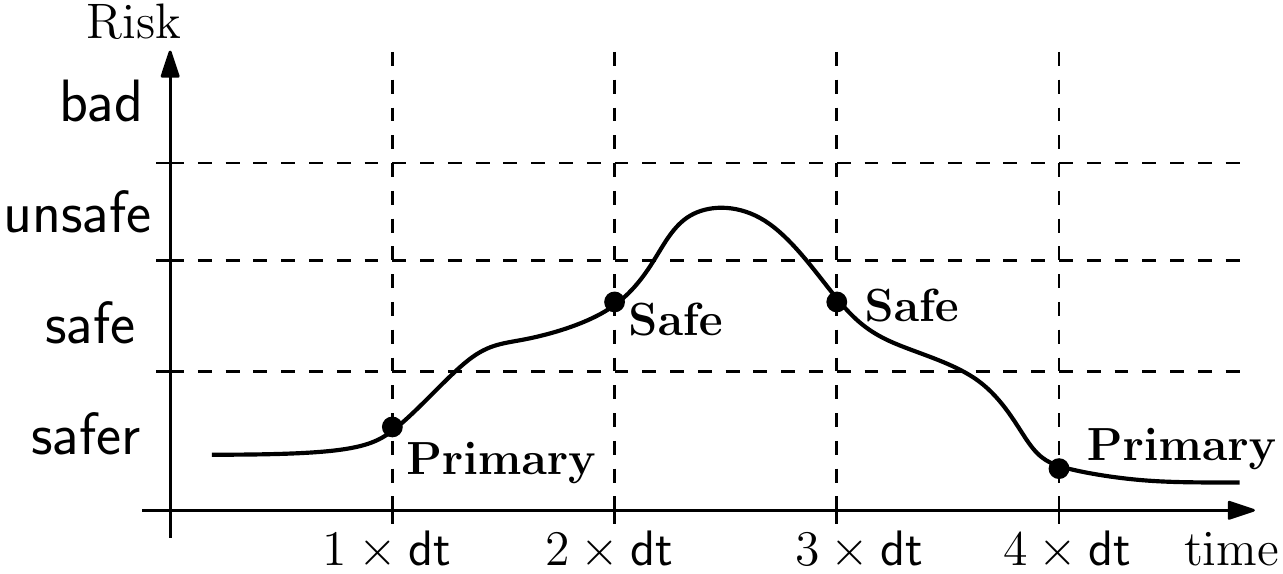}
  \caption{Illustration of how one expect \RAP to maintain safety during runtime. $\dt$ is the sampling time of the decision module. \textbf{Primary} (respectively, \textbf{Safe}) denotes that the decision module switches to the primary (respectively, safe) controller.}
  \label{fig:rta-dynamics}
  \vspace{-5mm}
\end{wrapfigure}

Runtime assurance architectures (\RAPs), based on the well-known simplex architecture~\cite{simplex,simplex2}, have been proposed~\cite{shankar,ramakrishna20jsa,mehmood22nfm} as a means to overcome this challenge.  
An \RAP contains a decision module that evaluates the system's safety risk formalized as a collection of safety properties. Whenever a safety risk is higher than acceptable, the \RAP moves the system to a safe state. 
\omitthis{
\begin{wrapfigure}{r}{0.6\textwidth}
    \includegraphics[width=0.58\textwidth]{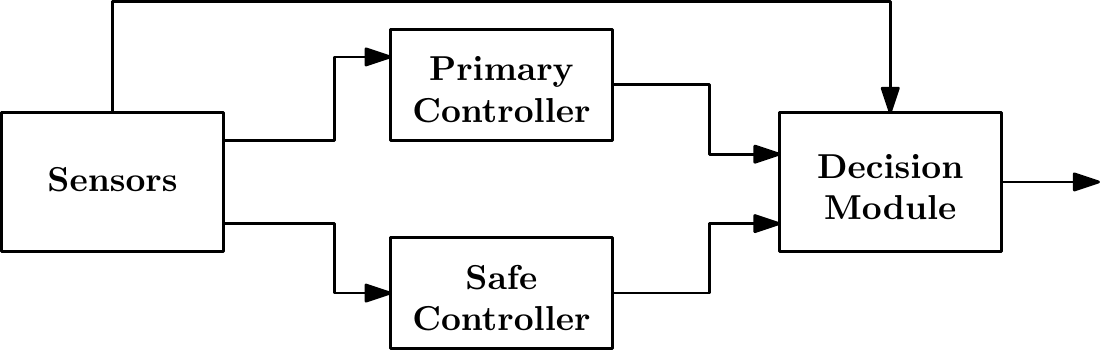}
  \caption{Simplex runtime assurance architecture.}
  \label{fig:simplex}
  \vspace{-7mm}
\end{wrapfigure}
The simplex architecture~\cite{simplex}, depicted in Figure~\ref{fig:simplex}, is an instance of an \RAP. 
Here, the decision module forwards the output of a primary, unverified controller whenever the system is in safe state; otherwise, if the system is at risk, the output of the safe, verified controller is used.
For example, a safe vehicle controller may decide to reduce vehicle speed to perform a Minimum Risk Maneuver (MRM) as suggested by current standards~\cite{unece}.
}
As illustrated by Figure~\ref{fig:rta-dynamics}, \RAP increases CPS safety and resilience by dynamically adapting the CPS behavior according to the perceived system risk level, recovering the CPS from a higher-risk situation.
We use the symbol \dt to denote the sampling interval in which the decision module evaluates the system's level of risk. 
These levels of risks are formalized as properties tailored according to the operational domain of the system~\cite{nigam22wrla}. For example, vehicles on a highway have a different formalization of risk level than vehicles in urban scenarios where pedestrians may be crossing roads. 
In the diagram in Figure~\ref{fig:rta-dynamics} there are four increasing levels of risk (\safer, \safe, \unsafe, \bad), \eg, denoting risks of an accident, from \safer\ denoting the lowest and desirable risk level to \bad\ denoting the highest level of risk that has to be avoided at all costs, to avoid possible accidents.
If the risk is \safer, then the decision module uses the output from the primary, unverified controller. However, if a higher risk \safe\ is detected, then the decision module uses the output of the safe controller.
The expectation is then that the safe controller recovers eventually from the high risk situation leading the system to return to a situation that is \safer. 
It may be that in the process the CPS will pass through situations that are \unsafe, but it definitely shall not pass through situations that are \bad, \eg, situations of imminent crash that trigger other safety mechanisms, such as emergency brakes.

There are two key properties about \RAPs which engineers have to demonstrate by providing sufficient evidence:
\begin{itemize}
  \item \textbf{\dt Adequacy:} the sampling time interval
  is small enough that bad situations are not missed by
  the \RAP;
  \item \textbf{Time Bounded Recoverability:} if the system 
  risk becomes greater than acceptable (\safer) the
  safe controller can bring the system back to a \safer\ state
  within a specified time bound, without entering a \bad\ state.
\end{itemize}

The main goal of this paper is to develop methods to generate formal proofs for these properties for \RAP instances in an automated fashion. This is accomplished by using the Symbolic Soft-Agents framework~\cite{nigam22wrla} which enables the automated generation of safety proofs for CPS using symbolic rewriting modulo SMT~\cite{rocha-meseguer-munoz-2017jlamp}. 
Our contributions here are in two areas: 
% formal foundation; and algorithms for checking  \RAP properties.
\begin{itemize}
\item \textbf{Formal foundation.}
We provide formal definitions for three variants of
\dt\ adequacy, and prove the relations among them.
We also provide a formal definition of time bounded
recoverability. We define a notion of one period recoverability, and prove
that one period recoverability together with any one of the
\dt\ adequacy properties implies time bounded
recoverability. The formal definitions are tailored so that they are amenable to automated verification.

\item \textbf{Automated Checking of \RAP Properties:} 
Based on the specification of \RAP properties and of abstract descriptions of situations in which CPSes operate, called logical scenarios~\cite{scenarios,menzel18iv}, we present algorithms for verifying two forms of \dt adequacy and for one period recoverability, and report
results of experiments for two logical scenarios. 
The experiments demonstrate the feasibility of automated
proof and also illustrate some of the challenges.
\end{itemize}

\noindent
% A key observation is that one can use the bounds on the dynamics of CPS agents, specified by the maximum speeds and accelerations, to enumerate all possible ways safety properties may change during execution. 
% This enumeration allows to check the adequacy of \dt. 
% A key insight is that if the safety properties are too coarse, then \dt\ adequacy requires very small \dt values, which would make it impractical. 
  % A solution is to consider a greater number of levels of safety.

\omitthis{  A second insight, also observed in our previous work~\cite{nigam22wrla},
 is the trade-offs between how much verification one delegates to the rewriting engine and how much to the SMT-solver. 
  We implemented the machinery for automated checking \RAP properties by taking these issues into account.

The case studies use very basic safe controllers to
minimize complexity of proof search while developing
the methods. This has the side benefit of illustrating
the subtlety of counter examples and the importance of
verifying resilience complementing basic reachability
search.
}
Section~\ref{sec:example} describes the logical scenarios of our running examples. Section~\ref{sec:safety_props} formalizes the notion of levels of risk using safety properties. These are then used to define several notions of sampling time adequacy in Section~\ref{sec:dt-adequacy} and recoverability properties in Section~\ref{subsec:resilience}. Section~\ref{sec:experiments} describes experiments based on the logical scenarios in Section~\ref{sec:example}. We conclude with related and future work in Sections~\ref{sec:related} and~\ref{sec:conc}. 
% Missing proofs and more details of the specification of the pedestrian crossing are shown in the Appendix.

\section{Logical Scenarios and Motivating Examples}
\label{sec:example}
%!TEX root = tase23.tex
\newcommand\at{\ensuremath{\mathtt{at}}\xspace}
\newcommand\dir{\ensuremath{\mathtt{dir}}\xspace}
\newcommand\ped{\ensuremath{\mathtt{p1}}\xspace}
\newcommand\Loc{\ensuremath{\mathtt{Loc}}\xspace}
\newcommand\KItem{\ensuremath{\mathtt{KItem}}\xspace}
\newcommand\RealSym{\ensuremath{\mathtt{RealSym}}\xspace}
\renewcommand\veh{\ensuremath{\mathtt{vh}}\xspace}

A key step in the development of autonomous CPSes is the definition of the situations in which these systems will operate~\cite{scenarios,menzel18iv,westhofen22metrics}. 
These situations are specified as abstract scenarios, called logical scenarios~\cite{scenarios,menzel18iv}, such as lane changing or vehicle following or pedestrian crossing, in which an AV has to avoid harm.
These logical scenarios contain details about the situations in which a vehicle shall be able to safely operate such as which types and number of actors, \eg, vehicles, pedestrians, operating assumptions, \eg, range of speeds, and road topology, \eg, number of lanes.\
Moreover, these logical scenarios are associated with safety metrics that formalize the properties that need to be satisfied by the vehicle. For a comprehensive list of logical scenarios and associated properties we refer to~\cite{westhofen22metrics} and references therein.  Examples of scenario description and
generation formalisms can be found in \cite{fremont19pldi,sifakis-bozga}.
As a logical scenario may have infinitely many concrete instances, it is challenging to demonstrate that a vehicle will satisfy such safety properties in all instances.

% An alternative, complementary approach is to provide  proofs that the \RAP instance satisfies the properties above~\cite{shankar}. 

\begin{figure}[t]
\begin{center}
  \includegraphics[width=0.47\textwidth]{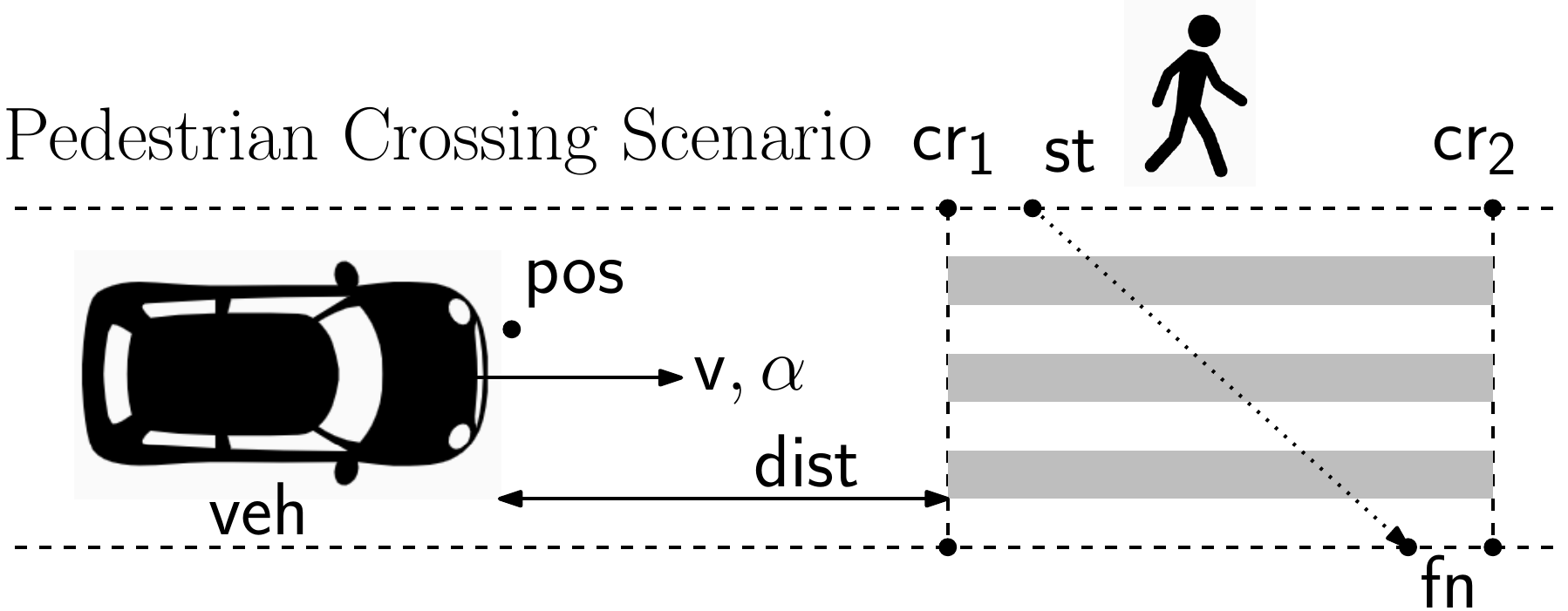} 
  \qquad
  \includegraphics[width=0.47\textwidth]{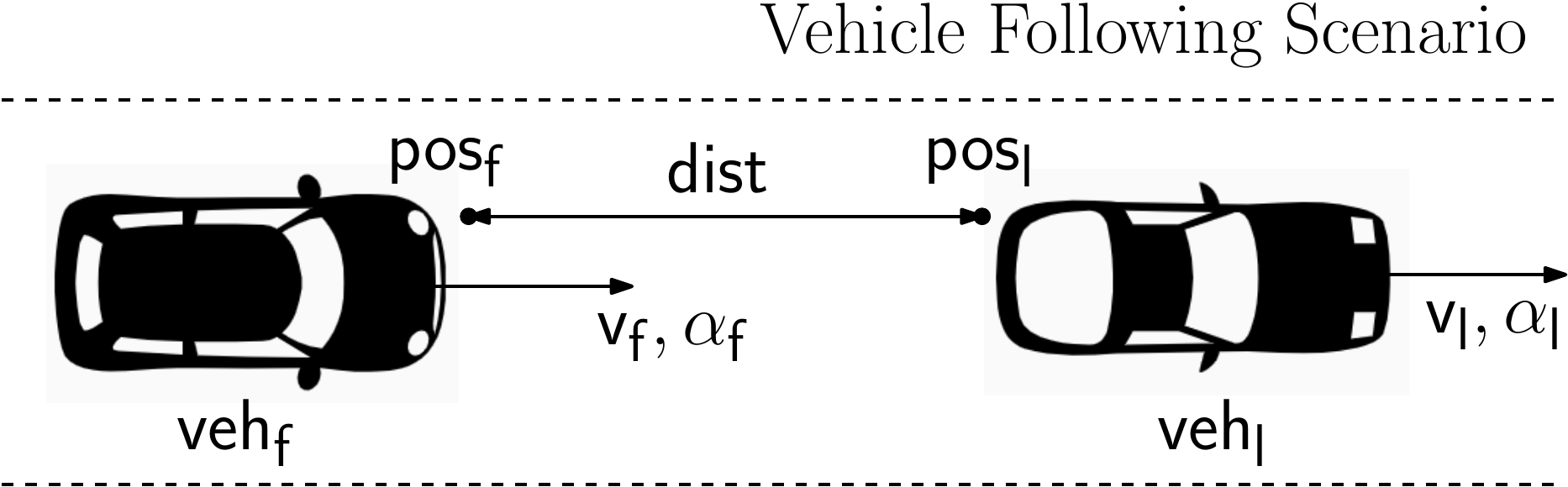}  
\end{center}
\vspace{-4mm}
\caption{Pedestrian crossing and vehicle following logical scenarios diagrams. The road is on the Y-axis, so imagine the illustrations rotated counterclockwise.}
\label{fig:pedestrian}
\vspace{-4mm}
\end{figure}

We use two running examples illustrated by the diagrams in Figure~\ref{fig:pedestrian}: a pedestrian crossing scenario and a vehicle following scenario.

\paragraph{Pedestrian Crossing}
In this scenario an ego vehicle~\footnote{An ego vehicle is a vehicle which is of primary interest in testing, trailing or operational scenarios.}, 
\veh, is at position $\pos$ and is approaching with speed $\vel$ and acceleration $\acc$, with a pedestrian crossing situated between the positions $\crossSt$ and $\crossFn$. 
Moreover, a pedestrian is attempting to cross the road using the pedestrian crossing. 
As long as the pedestrian does not move outside the pedestrian crossing, the exact shape of the pedestrian crossing is not important as \veh shall always stop before the pedestrian crossing whenever a pedestrian is intending to cross it.
To keep things simple, assume that the pedestrian is crossing the street at constant speed, \pvel, following a straight line as illustrated in Figure~\ref{fig:pedestrian} by the dashed line from $\pedA$ to $\pedB$. 

The operational design domain (ODD) of such a logical scenario is specified by constraints on its parameters (\pos, \vel, \acc, \crossSt, \crossFn, \pvel). Typically, one specifies the bounds on the speeds and accelerations. Consider for example:

\(
  \begin{array}{c}
   0 m/s \leq \vel \leq 10 m/s \quad -8 m/s^2 \leq \acc \leq 2 m/s^2 \quad 1 m/s \leq \pvel \leq 4 m/s
  \end{array}
\)

\noindent
Moreover, $\pos.y < \crossSt.y$, that is the vehicle is approaching the pedestrian crossing and  $\crossSt.y \leq \pedA.y, \pedB.y \leq \crossFn.y$, that is  $\pedA, \pedB$ are in the pedestrian crossing area, where for any position $\mathsf{l} = (\px,\py)$, $\mathsf{l}.x$ and $\mathsf{l}.y$ denote, respectively, $\px$ and $\py$.

\paragraph{Vehicle Following} 
Our second running example is a vehicle following scenario as depicted in Figure~\ref{fig:pedestrian}. 
This example commonly appears in the literature and therefore, we do not describe in the same level of detail, but simply refer to~\cite{nigam22wrla}. In a nutshell, it consists of two vehicles, a follower vehicle (\vehf) and a leader vehicle (\vehl). Typically, these vehicles are in a highway with multiple lanes at reasonably high speeds, \eg, speeds between $60 km/h$ and $140km/h$ and the same acceleration bounds as in the vehicle in the pedestrian crossing scenario. Moreover, there are only vehicles, \ie, no pedestrians. no bicycles, etc. 
The following vehicle shall avoid approaching dangerously close to the leader vehicle while still maintaining a reasonable speed.
% This scenario is described in more detail in~\cite{nigam22wrla}.

\smallskip

% Notice that a logical scenario may have infinitely many instances. Therefore, producing evidence through simulation or testing requires a very high number of simulations and even then it is possible that some edge-cases are missed for which the \RAP instance does not satisfy the properties described above.

% The task of the safety engineer is to provide evidence that the vehicle implementation will behave safely for all logical scenario instances.  
% This is even more daunting as sensors may only be able, for example, to detect a pedestrian within some range or fail to detect the exact position of pedestrian crossing due to bad visibility of the camera used by the sensor. 
% Such issues has motivated the standard Safety Of The Intended Functionality (SOTIF)~\cite{sotif} for safety of vehicles that operate with less human intervention.

We assume that from an instance, \conf, of a logical scenario (\LS), we can compute the function $\conf \lra_\Delta \conf_1$, where $\conf_1$ is an $\LS$ instance specifying the physical attributes, \eg, speeds, directions, accelerations, of the agents obtained according to their speeds, directions and accelerations in \conf after a period of $\Delta > 0$ time units. 
Moreover, we assume that if $\conf \lra_{\Delta_1 + \Delta_2} \conf_1$, then there exists $\conf'$ such that $\conf \lra_{\Delta_1} \conf'  \lra_{\Delta_2} \conf_1$.
\footnote{Since $\lra_\Delta$ is a function,  \conf' is unique.}
% \red{During this time we assume that the acceleration is constant}
For example, consider the instance of the pedestrian crossing scenario where the vehicle has speed of $10m/s$,  acceleration of $2m/s^2$, and position $\pos.x = 0m$. 
After $\Delta = 0.1s$, the speed of the vehicle will be $10.2m/s$ and new position $1.1m$. The vehicle is traveling
in a constant direction along the road, an we omit it.

% We show how this relation can be specified using executable symbolic specifications in Section~\ref{sec:executable-semantics}.

% We use this example to illustrate how one can execute logical scenarios (instead of their instances) using the Symbolic Soft Agents framework proposed in our previous work~\cite{nigam22wrla}. We refer to that paper for more details about the framework.

\section{Safety Properties and Levels of Risk}
\label{sec:safety_props}
\label{subsec:safety-properties}
%!TEX root = tase23.tex
A key aspect of \RAP mechanisms is the ability to check for the level of risk of the system, \eg, whether it is safe or not. 
We formalize the notion of level of risk as a partial order on safety properties as follows: 
\begin{definition}
  An \RAP safety property specification for a logical specification $\LS$ is a tuple $\tup{\Sscr,\prec_1,\bad,\vDash}$ where 
\begin{itemize}
  \item $\Sscr = \{\SP_1, \ldots, \SP_n\}$ is a finite set of safety properties;
  \item $\prec_1 : \Sscr \times \Sscr$ is an asymmetric binary relation over $\Sscr$, where $\SP_1 \prec_1 \SP_2$ denotes that the safety property $\SP_2$  specifies a less risky condition than the safety property $\SP_1$. Let $\prec$ be the order obtained from $\prec_1$ by applying transitivity. We assume that $\prec$ is a strict pre-order
  (no cycles).
  \item the safety property $\bad \in \Sscr$ is the least element of $\prec$, specifying the condition that shall be avoided, \ie, the highest risk
  \item $\vDash$ specifies when an instance \conf of $\LS$ satisfies a property $\SP \in \Sscr$, written $\conf \vDash \SP$. 
  Moreover, we assume that if $\conf \vDash \SP_1$ and $\SP_1 \prec \SP_2$ or $\SP_2 \prec \SP_1$, then $\conf \nvDash \SP_2$. That is, any instance of a logical scenario can only satisfy one level of risk. We also assume that any instance of a logical scenario is at some level of risk, that is, for all instances $\conf$ of \LS, there is at least one $\SP$ such that $\conf \vDash \SP$.
\end{itemize}
\end{definition}

The following two examples illustrate different options of safety properties for the pedestrian crossing and the vehicle following examples described in Section~\ref{sec:example}. 

\begin{example}
\label{ex:sps-pedestrian}Consider the pedestrian crossing shown in Figure~\ref{fig:pedestrian}. 
We define the following \RAP safety property specification  $\tup{\{\bad,\unsafe,\safe,\safer\},\prec_1,\bad,\vDash}$ with $\bad \prec_1 \unsafe \prec_1 \safe \prec_1 \safer$ based on the \emph{time to zebra} metric~\cite{westhofen22metrics}
\footnote{Zebra is the pedestrian crossing zone.}
\begin{equation}
\label{eq:ped_only_relative}
 \begin{array}{lcl}
\safer & := & \dist \geq \dstop + \gapsafer * \vel\\
\safe & := & \dstop + \gapsafer * \vel  > \dist \geq \dstop + \gapsafe * \vel \\
\unsafe & := & \dstop + \gapsafe * \vel  > \dist \geq \dstop + \gapunsafe * \vel \\
\bad & := & \dstop + \gapunsafe * \vel  > \dist\\
\end{array}   
\end{equation}
where $\dist = \crossSt.y - \pos.y$ is the distance between the ego vehicle
and the pedestrian crossing,
 $\dstop = - (\vel * \vel) / (2 * \maxDec)$ is the distance necessary to stop the ego vehicle by applying its maximum deceleration \maxDec, \eg, when issuing an emergency brake, and $\gapsafer > \gapsafe > \gapunsafe > 0$ are used with \vel\ to specify a safety margin distance in the safety property. 
The values for $\gapsafer,\gapsafe,\gapunsafe$ shall be defined according to the ego vehicle's capabilities, \eg, the sampling time \dt, and the ODD specifications, \eg, bounds on acceleration and speed. 
It is then straightforward to check whether an instance of a pedestrian logical scenario satisfies ($\vDash$) any one of the properties above.

While this may seem like a good candidate safety property specification for the pedestrian crossing, it turns out that it is hard to demonstrate vehicle recoverability as we show in Section~\ref{sec:experiments}. 
The problem lies in the fact that the three properties tend to be all the same when the vehicle speed (\vel) tends to zero, and similarly, when $\dist$ is too large.
We, therefore, establish an alternative definition for $\safer$ as follows:
\begin{equation}
\label{eq:sp_ped_bounds}
\begin{array}{lcl}
\safer & := & \dist \geq \dstop + \gapsafer * \vel \textrm{ or } \vel \leq \lowSpd \textrm{ or } \dist \geq \farAway 
\end{array}   
\end{equation}
where $\lowSpd$ and $\farAway$ are constants specifying a maximum speed for which the vehicle is very safe, \eg, the speed $\lowSpd$ is less than the speed of a pedestrian, and the distance $\farAway$ that is far enough from the pedestrian crossing.
\end{example}

% ( ((reactTime + dt + dt) * v1) + (1/2 * (reactTime + dt + dt) * (reactTime + dt + dt) * maxacc1) -
%                ((v1 + (maxacc1 * (reactTime + dt + dt))) * (v1 + (maxacc1 * (reactTime + dt + dt))) / (2/1 * maxdec1)) +
%                ((v0 * v0) / (2/1 * maxdec0)) ))

\begin{example}
\label{ex:rss}
One well-known example for vehicle safety assurance for the vehicle following scenario
is the Responsibility-Sensitive Safety (RSS)~\cite{rss,westhofen22metrics} safe distance metric. The RSS safety distance $\drss(\react)$ is specified as follows:

\(
  \drss(\react) = \vel \times \react + \frac{\maxacc_f \times \react^2}{2} - \frac{(\vel + \maxacc_f \times \react)^2}{2\times\maxdec_f} - \frac{\vel_l^2}{2 \times \maxdec_l}
\)

\noindent
where $\react$ is a parameter for the time for the vehicle to react; $\vel$ and $\vel_l$ are, respectively, the follower and leader vehicle speeds; $\maxacc_f$ is the maximum acceleration of the follower vehicle; and $\maxdec_f$ and $\maxdec_l$ are, respectively, the maximum deceleration of the follower and leader vehicles.
Based on $\drss(\react)$ two properties are defined:
$\bad$ when $\dis < \drss$ and $\safer$ otherwise. 

% In~\cite{rss} requirements for behavior of both vehicles
% when the $\bad $- $\safer$ threshold is crossed
% are defined and inductive arguments are given 
% for different road and traffic configurations showing that
% if all vehicles meet the requirements then there will be
% no collisions.    

As RSS has only two properties, the definition of recoverability using \RAP implies that the system must always satisfy
the $\safer$ property; otherwise it must satisfy $\bad$. 
This means that the primary controller shall be trusted and that \RAP is not necessary from the beginning (and probably not desired as the primary controller is assumed not to be verified).
It is possible to adapt the RSS definitions by adding additional levels in between \safer\  and \bad\ based on the $\react$ time:
\[
\begin{array}{l@{\quad}l}
  \safer := \dis \geq \drss(3 \times \dt) & 
  \safe := \drss(2 \times \dt) \leq \dis < \drss(3 \times \dt)\\
  \unsafe := \drss(\dt) \leq \dis < \drss(2 \times \dt) & 
  \bad := \dis < \drss(\dt)
\end{array}
\]
Intuitively, when a vehicle is in a configuration satisfying $\safer$ it can wrongly evaluate safety risk, \eg, due to distance sensor errors, for two cycles before the RSS property is invalidated. 
Similarly, $\safe$ it can evaluate wrongly for one cycle and $\unsafe$ it always has to evaluate correctly the risk.
\end{example}

\section{Sampling Time (\dt) Adequacy}
\label{sec:dt-adequacy}
%!TEX root = tase23.tex

The \RAP monitor has to detect when the system risk changes, and even more so when risk increases, that is, when systems satisfy properties $\SP$ that are closer to \bad, \ie, move lower in the order $\prec$. This means that the sampling time \dt plays an important role in the correctness of a \RAP system. 
For example, if the sampling time is $4 \times \dt$ in Figure~\ref{fig:rta-dynamics}, the \RAP monitor may fail to detect elevation of risk from \safer\ to \safe\ thus not activating the trusted controller soon enough to avoid further escalation of risk.

% \begin{wrapfigure}{r}{0.6\textwidth}
% \vspace{-8mm}
%     \includegraphics[width=0.6\textwidth]{figures/rta-bad-dt.pdf}
%   \caption{A large \dt may not be able to identify risky situations and therefore, fail to activate the trusted controller.}
%   \label{fig:rta-bad-dt}
%   \vspace{-7mm}
% \end{wrapfigure}

There is a trade-off between the ability of the system to detect changes of risk and therefore its ability to quickly react to changes, and the performance requirements of monitor system in determining risk, \ie, \dt time. 
The lower the \dt, the greater is the ability of the system to detect changes and also greater are the performance requirements on the monitoring components.

Moreover, a key challenge is that \dt shall be appropriate in detecting risk changes for all instances of the ODD, \ie, all possible instances of speeds and accelerations.
Our approach is to use SMT-solvers to generate \dt adequacy proofs automatically building on ideas in~\cite{nigam22wrla}.  Depending on the definition of \dt adequacy, the complexity of the problem can increase substantially, making automation  difficult or not feasible. 

\begin{figure}[t]
\begin{center}
  \includegraphics[width=0.99\textwidth]{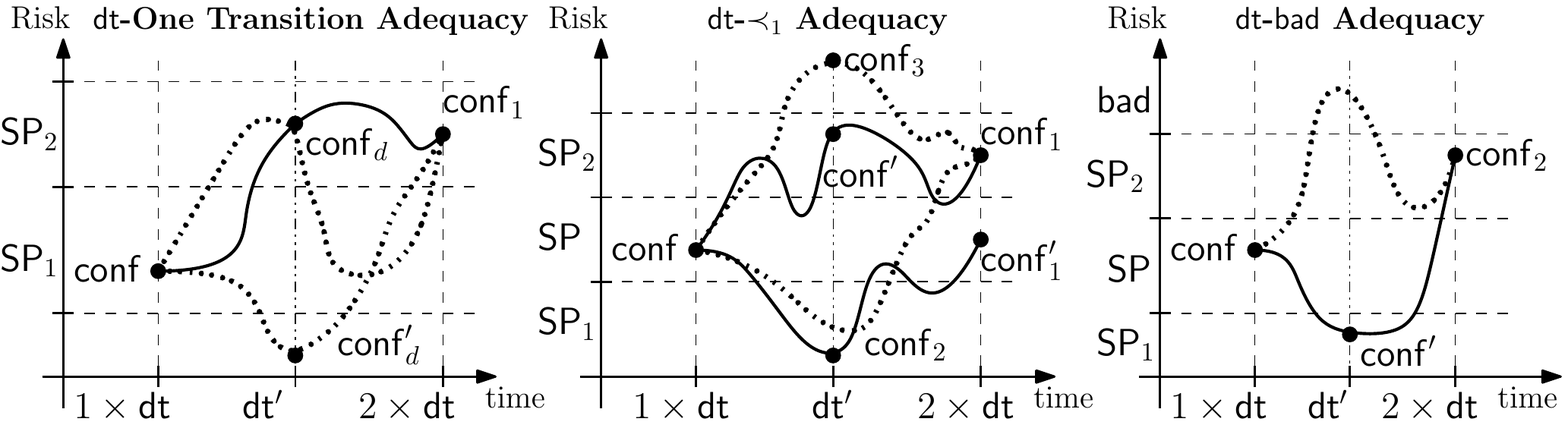}
\end{center}
\vspace{-5mm}
\caption{Illustration of \dt adequacy properties. Full line system evolutions illustrate allowed evolutions and dotted evolutions illustrate not allowed evolutions.}
\label{fig:dt-props}
\end{figure}

We propose three alternative definitions of requirements on \dt, defined below, that illustrate the trade-offs between the capability of the system to detect risk changes and the development and verification efforts. 
Figure~\ref{fig:dt-props} illustrates these definitions.
The first definition, called \emph{one transition adequacy}, is illustrated by left-most diagram in Figure~\ref{fig:dt-props}.
Intuitively, this definition states that the \dt shall be fine enough to detect whenever the configuration of the scenario evolves from
satisfying a property, $\SP_1$, to satisfying another property, $\SP_2$.
As an example, the dotted evolution of the system passing through $\conf_d'$ contains multiple property changes within a period of \dt. 
% If this is possible, the \dt is not one transition adequate and therefore, smaller sampling times shall be used.

\begin{definition}
\label{def:dt-one-transition}
  Let $Spec = \tup{\Sscr,\prec_1,\bad,\vDash}$ be a \RAP safety property specification for a logical scenario $\LS$; and \dt be a sampling time.
   \dt is \emph{one transition adequate} with respect to $Spec$ and \LS if for all instances $\conf$, $\conf_1$ of \LS such that $\conf \to_\dt \conf_1$ we have:
   \begin{itemize}
     \item if $\conf \vDash \SP_1$ and $\conf_1 \vDash \SP_2$, then there is a decomposition $\conf \to_{\dt'} \conf_d \to_{\dt - \dt'} \conf_1$ of $\conf \to_\dt \conf_1$ for some $0 \leq \dt' < \dt$, such that:
     \begin{itemize}
       \item For all decompositions of $\conf \to_{\dt'} \conf_d$ as $\conf \to_{\dt_2} \conf_2 \to_{\dt' - \dt_2} \conf_d$ where $0 < \dt_2 < \dt'$, we have that $\conf_2 \vDash \SP_1$;
       \item For all decompositions of $\conf_d \to_{\dt - \dt'} \conf_1$ as $\conf_d \to_{\dt_3} \conf_3 \to_{\dt - \dt' - \dt_3} \conf_1$ where $0 \leq \dt_3 < \dt-\dt'$, we have that $\conf_3 \vDash \SP_2$.
     \end{itemize}
   \end{itemize}
\end{definition}

The following proposition follows immediately from Definition~\ref{def:dt-one-transition}. It states that if
$\dt$ is one transition adequate, then to check that a 
configuration satisfying $\bad$ is not reachable, it is enough to check whether the configurations during sampling are not $\bad$, instead of checking all decompositions. 

\begin{proposition}
\label{prop:dt-one-transition}
Let $Spec = \tup{\Sscr,\prec_1,\bad,\vDash}$ be a \RAP safety property specification for a logical scenario $\LS$. Let $\dt$ be one-transition-adequate w.r.t. $Spec$. For all decompositions $\conf \rightarrow_{\dt'} \conf' \rightarrow_{\dt - \dt'} \conf_1$ of $\LS$, $\conf' \vDash \bad$ if and only if $\conf \vDash \bad$ or $\conf_1 \vDash \bad$. 
\end{proposition}

Definition~\ref{def:dt-one-transition} is rather complex involving many quantifier alternations thus being very difficult to generate proofs for. 
In fact, due to limitations on computing time, it is not always possible to guarantee that $\dt$ can satisfy one-transition-adequacy. Therefore, we propose 
two alternative definitions of weaker properties illustrated by the center and right-most diagrams in Figure~\ref{fig:dt-props}.
These properties are amenable to the automated generation of proofs as we detail in Section~\ref{sec:experiments}.

The first alternative definition is $\prec_1$ adequacy. Instead of requiring $\dt$ to be fine enough to detect 
when the system satisfies different properties, 
$\prec_1$ adequacy allows system evolution to migrate within $\prec_1$ range of a safety property multiple times, as illustrated by the system evolution passing through $\conf'$. 
The system shall be able to detect whenever the risk of the system increases at least two levels.
% It requires, however, the detect of whenever the system evolves at least  two levels higher than the $\prec_1$ risk range. This is illustrated by the evolution passing through $\conf_3$ as the system evolves to a property outside the $\SP_1 \prec_1 \SP \prec_1 \SP_2$ range within a period of $\dt$.

\begin{definition}
  Let $Spec = \tup{\Sscr,\prec_1,\bad,\vDash}$ be a \RAP safety property specification for a logical scenario $\LS$; and \dt be a sampling time.
   \dt is \emph{$\prec_1$ adequate} with respect to $Spec$ and \LS if for all instances $\conf$ of \LS and  relations $\conf \rightarrow_\dt \conf_1$ if for all $0 < \dt'\leq \dt$ and decompositions $\conf \rightarrow_{\dt'} \conf' \rightarrow_{\dt - \dt'} \conf_1$ we have:
   \begin{itemize}
     \item If $\conf \vDash \SP_1$ and $\conf_1 \vDash \SP_2$ for $\SP_1 \neq \SP_2$, then $\conf' \vDash \SP_1$ or $\conf' \vDash \SP_2 $.
     \item If $\conf \vDash \SP$ and $\conf_1 \vDash \SP$, then $\conf' \vDash \SP$ or $\conf' \vDash \SP'$ where $\SP' \prec_1 \SP$ or $\SP \prec \SP'$.
   \end{itemize}
\end{definition}

One can generalize the definition of $\prec_1$ to allow evolutions on larger ranges of safety properties, \eg $\prec_n$ adequacy for $n \geq 1$ allow evolutions within $n$ safety risk levels.

The following property of $\prec_1$-adequacy provides a basis for defining recoverability based on $\prec_1$-adequate \dt. It is enough to check that no configuration satisfying $\bad$ or a property immediately greater to $\bad$ is reachable.

\begin{proposition}
\label{prop:prec1}
  Let $Spec = \tup{\Sscr,\prec_1,\bad,\vDash}$ be a \RAP safety property 
  specification for a logical scenario $\LS$; and \dt be $\prec_1$ adequate 
  sampling time.  If $\conf \to_\dt \conf_1$ with $\conf \vDash \SP_1$ 
  and  $\conf_1 \vDash \SP_2$ where  $\SP_1 \neq \bad$ and  $\SP_2 \neq \bad$ and  $\bad \nprec_1 \SP_1$ or $\bad \nprec_1 \SP_2$,  then for all $0 < \dt'\leq \dt$ and decompositions $\conf \rightarrow_{\dt'} \conf' \rightarrow_{\dt - \dt'} \conf_1$ we have $\conf' \nvDash \bad$.
\end{proposition}

\omitthis{
\begin{proof}
  Assume that $\SP_1 \neq \SP_2$. 
  In this case, from the $\prec_1$ property, we have $\conf' \vDash \SP_1$ or $\conf_1 \vDash \bad$ since $\SP_1 \neq \bad$ and $\SP_2 \neq \bad$. 
  Assume that $\SP_1 = \SP_2 = \SP$, with $\bad \nprec_1 \SP$ (from the assumption $\bad \nprec_1 \SP_1$ or $\bad \nprec_1 \SP_2$. 
  Then $\conf_1 \vDash \SP'$ where  $\SP' = \SP$, in which case $\conf \nvDash \bad$, $\SP \prec_1 \SP'$, in which case from $\bad \nprec_1 \SP$, $\conf \nvDash \bad$, or $\SP \prec_1 \SP'$, in which case, from the fact that $\bad$ is the least element in the pre-order, $\conf \nvDash \bad$.
\end{proof}
}

Consider for example the safety property specification in Example~\ref{ex:sps-pedestrian} and assume that $\dt$ is $\prec_1$-adequate. From Proposition~\ref{prop:prec1}, if there is no transition $\conf \to_\dt \conf_1$ where $\conf \vDash \unsafe$ and $\conf_1 \vDash \unsafe$, then we can guarantee that the system does not pass through a configuration $\conf'$ with $\conf' \vDash \bad$ including the intermediate configurations that have not been sampled by the vehicle system.

The next adequacy only requires that the \dt is fine enough to detect when a system evolution satisfies the $\bad$ property. 
As illustrated by the right-most diagram in Figure~\ref{fig:dt-props}, the dotted evolution satisfying $\bad$ within $\dt$ would invalidate $\dt$ adequacy.

\begin{definition}
  Let $Spec = \tup{\Sscr,\prec_1,\bad,\vDash}$ be a \RAP safety property specification for a logical scenario $\LS$; and \dt be a sampling time.
   \dt is \emph{$\bad$-adequate} with respect to $Spec$ and \LS if for all instances $\conf$ of \LS and  $\conf \rightarrow_\dt \conf_1$ if for all $0 < \dt'\leq \dt$ and decompositions $\conf \rightarrow_{\dt'} \conf' \rightarrow_{\dt - \dt'} \conf_1$ we have:
   \begin{itemize}
     \item if $\conf \nvDash \SP$ and $\conf_1 \nvDash \SP$ with $\SP = \bad$ or $\bad \prec_1 \SP$, then $\conf' \nvDash \bad$.
   \end{itemize}
\end{definition}

The following proposition is similar to Proposition~\ref{prop:dt-one-transition} establishing the conditions for verifying for $\bad$-adequacy.

\begin{proposition}
\label{prop:dt-bad}
Let $Spec = \tup{\Sscr,\prec_1,\bad,\vDash}$ be a \RAP safety property specification for a logical scenario $\LS$. Let $\dt$ be $\bad$-adequate w.r.t. $Spec$. For all decompositions $\conf \rightarrow_{\dt'} \conf' \rightarrow_{\dt - \dt'} \conf_1$ of $\LS$, $\conf' \vDash \bad$ if and only if $\conf \vDash \SP_0$ and $\conf_1 \vDash \SP_1$ with $\{\SP_0, \SP_1\} \subseteq \{\bad\} \cup \{\SP \mid \bad \prec_1 \SP\}$. 
\end{proposition}

The following proposition establishes relations between the different adequacy definitions. Our experiments show that it is possible for $\dt$ to be $\bad$-adequate and not $\prec_1$-adequate. 

\begin{proposition}
\label{prop:bad-dt-adeq}
  Let $Spec = \tup{\Sscr,\prec_1,\bad,\vDash}$ be a safety property specification for a logic scenario \LS and $\dt$ a sampling time.
  \begin{itemize}
    \item If \dt is one transition adequate, then \dt is $\prec_1$-adequate and $\dt$ is \bad-adequate.
    \item If \dt is $\prec_1$-adequate then $\dt$ is \bad-adequate.
  \end{itemize} 
\end{proposition}

% \red{Explain the procedure of checking adequacy. I think it is ok. We explain above how the procedure looks like.}

% \paragraph{Procedure for Checking Time Sampling Adequacy}
% 1. Enumerate all possible conf -> conf' such that 

\section{\RAP-based Recoverability Properties}
\label{sec:rta-prop}
\label{subsec:resilience}
%!TEX root = tase23.tex
There are many informal definitions of resilience~\cite{allenby2005toward,barker2013resilience,bloomfield2020towards,laprie2008dependability}. 
In the broadest sense, resilience is  
``the ability of a system to adapt and respond to changes (both in the environment and internal)''~\cite{bloomfield2020towards}.
NIST~\cite{ross2019developing} provides a more precise, but still informal definition of resilience and more focused on attacks: ``The ability to anticipate, withstand, recover, and adapt to adverse conditions, stresses, attacks or compromises on systems that use or are enabled by cyber resources.''

Intuitively, systems, such as an autonomous vehicle in an \LS instance, implementing  \RAP can be shown to
exhibit a basic form of resilience we refer to as
recoverability:  they detect when a specified risk level
is reached and adapt to reduce the risk.
Our goal is to formalize this intuition of \RAP recoverability with precise definitions.

\omitthis{CLT:  need to state that -->_dt is a function
(functional relation?) since notation has changed.
Maybe say that conf has two parts logical (agents
knowledge) and physical (the physical state).
tasks changes the logical, dt changes the physical
part.
}

To accomplish this, we must model the control aspects
of the system as well as the change in physical state.
Thus we augment the semantic relation $\to_\dt$ 
(called $\lra_\Delta$ above) 
which models the
physical aspect of behavior 
with a  relation $\to_\tasks$ that
models the control aspect, typically sensing, analyzing, and deciding/planning. Formally, the system behavior is a set of (possibly infinite) execution traces:

\(
  \conf_0 \to_\tasks \conf_0' \to_{\dt} \conf_1 \to_\tasks \conf_1' \to_{\dt} \conf_2 \to_\tasks \cdots
\)

\noindent
where $\dt$ is the system's sampling time, 
$\conf_{i}' \to_\dt \conf_{i+1}$ is a function,
and $\conf_{i}\to_\tasks \conf_i'$ is an internal transition specifying the behavior of the agents in $\conf_i$, \eg, sensing, updating local knowledge bases, and deciding which actions to take. The exact definition of this transition depends on system specification. 
% In Section~\ref{sec:executable-semantics}, we demonstrate how to specify it symbolically.
Since safety properties are related to the physical attributes of the system, \eg, speed, location, we normally assume that if $\conf_i \vDash \SP$, then also $\conf_{i}' \vDash \SP$. For example, this is the case with the safety properties in Example~\ref{ex:sps-pedestrian}.
This assumption is not strictly necessary as the definitions below can be extended to cover cases when this assumption does not hold.

\begin{definition}
\label{def:resilience}
  Let $Spec = \tup{\Sscr,\prec_1,\bad,\vDash}$ be a safety property specification for a logical scenario \LS and $\dt$ a sampling time, where $\SP_\safe \in \Sscr$ is the minimal acceptable safe property and $\SP_\safer \in \Sscr$ is the acceptable safer property where $\SP_\safe \prec \SP_\safer$.
  Let $t$ be a positive natural number.
  A system $S$ is $\tup{\SP_\safe,\SP_\safer,t}$-recoverable if for all instances $\conf_0$ of $\LS$ and traces $\tau = \conf_0 \to_\tasks \conf_0' \to_{\dt} \conf_1 \to_\tasks \conf_1' \to_{\dt} \cdots$ such that $\conf_0 \vDash \SP$ with $\SP = \SP_\safe$ or $\SP_\safe \prec \SP$:
  \begin{itemize}
    \item For all $\conf_{i}' \to_\dt \conf_{i+1}$ in $\tau$, there is no decomposition $\conf_{i}' \to_{\dt_1} \conf \to_{\dt - \dt_1} \conf_{i+1}$, with $0 \leq \dt_1 \leq \dt$, such that $\conf \vDash \bad$. That is, the system never reaches a configuration that satisfies $\bad$.
    \item For all $\conf_i$ in $\tau$, such that $\conf_i \vDash \SP_\safe$, then $\conf_{i + t} \vDash \SP$ with $\SP_\safer \prec \SP$ or $\SP_\safer = \SP$. That is, if the system reaches the minimal safety property, it necessarily returns to the acceptable safer property.
  \end{itemize}
\end{definition}

This definition formalizes the ability of the system to recover from a higher level of risk as illustrated by Figure~\ref{fig:rta-dynamics}.
Intuitively, the property $\SP_\safe$ specifies the highest acceptable risk before the system shall react to reduce risk, \ie, when the \RAP instance triggers the safe controller, while $\SP_\safer$ specifies the risk that shall be achieved within $t$ logical ticks of the system, \ie, $t \times \dt$, that is when the \RAP instance resumes using the output of the primary controller. 

There are some subtleties in this definition that are worth pointing out:
\begin{itemize}
  \item Recovery Period: The time $t$ in Definition~\ref{def:resilience} specifies the time that the system has to recover. On the one hand, it avoids that the system stays in a higher risk situation, albeit still safe, for a long period of time, thus reducing the chance of safety accidents. On the other hand, if $t$ is too small, it will require a stricter safe controller or not be realizable given the vehicle's capabilities, \eg, maximum deceleration.
  Therefore, the value of $t$ will depend on situation under consideration. To mitigate this problem, we propose automated ways to prove recoverability in Section~\ref{sec:experiments}.
  
  \item Recoverability Smoothness: Notice that we require that $\SP_\safe \prec \SP_\safer$ and not $\SP_\safe \prec_1 \SP_\safer$, \ie, $\SP_\safer$ can be multiple levels of risk safer than $\SP_\safe$. 
  By selecting appropriately these properties, \eg, setting $\SP_\safer$ with a much lower risk than $\SP_\safe$, one can avoid the oscillation of  the system between normal operation (using the primary controller) and recovery operation (using the safe controller). 
\end{itemize}

\paragraph{Procedure to Demonstrate Recoverability}
A challenge in proving a system resilient as per Definition~\ref{def:resilience} is that one needs to reason about all traces which may have infinite length and furthermore all decomposition of traces. 
To address this challenge we demonstrate (Theorem~\ref{th:resilience} below) that it is enough that the $\dt$ is adequate (as in Section~\ref{subsec:safety-properties}), $\dt$ is fine enough not to skip properties (Definition~\ref{def:skip} below), and consider only traces of bounded size as specified by the following definition:

\begin{definition}
\label{def:one-period-resilience}
  Let $Spec, \LS, \SP_\safer, \SP_\safe, t$ be as in Definition~\ref{def:resilience} and $\dt$ be the sampling time.
  A system $S$ is $\tup{\SP_\safe,\SP_\safer,t}-$one-period-recoverable if for all traces
  $\tau = \conf_0 \to_\tasks \conf_0' \to_{\dt} \conf_1 \to_\tasks \cdots \to_{\dt} \conf_{t}$ such that $\conf_0 \vDash \SP_\safe$:
  \begin{enumerate}
    \item $\conf_{t} \vDash \SP_\safer$--the system recovers in $t$ time ticks to a lower risk situation.
    \item \label{it:one-recovery-2} For all $\conf_{i}' \to_\dt \conf_{i+1}$ in $\tau$, there is no decomposition $\conf_{i}' \to_{\dt_1} \conf \to_{\dt - \dt_1} \conf_{i+1}$, with $0 \leq \dt_1 \leq \dt$, such that $\conf \vDash \bad$.
  \end{enumerate}
\end{definition}

To prove recoverability for unbounded traces (Theorem~\ref{th:resilience}), we also need to ensure that property $\SP_\safe$ that triggers an \RAP is not skipped. This is formalized by the following definition.

\begin{definition}
\label{def:skip}
Let $Spec, \LS, \SP_\safer, \SP_\safe, t$ be as in Definition~\ref{def:resilience} and $\dt$ be the sampling time. We say that $\dt$ does not skip a property $\SP_\safe$ if there is no transition of the form $\conf \lra_\dt \conf_1$ such that $\conf \vDash \SP$ and $\conf_1 \vDash \SP_1$ with $\SP_\safe \prec \SP$ and $\SP_1 \prec \SP_\safe$. 
\end{definition}

\begin{theorem}
\label{th:resilience}
Let $\dt$ be one-transition or $\prec_1$ or $\bad$-adequate
where  $\dt$ does not skip $\SP_\safe$. 
 A system $S$ is $\tup{\SP_\safe,\SP_\safer,t}$-one-period-recoverable  if and only if $S$ is $\tup{\SP_\safe,\SP_\safer,t}$-recoverable.
\end{theorem}

\paragraph{Condition for checking one-recovery-period recoverability:}
Even when considering only one-recovery-period recoverability, it is still necessary to consider all possible decompositions of $\dt$ transitions (item~\ref{it:one-recovery-2} in Definition~\ref{def:one-period-resilience}). 
This can be overcome depending on the type of $\dt$ adequacy:
using Propositions~\ref{prop:dt-one-transition},~\ref{prop:prec1}, and~\ref{prop:dt-bad},  it is enough to check that that all configurations $\conf_i$ for $0 \leq i \leq t$ do not satisfy $\bad$ nor a $\SP$ such that $\bad \prec_1 \SP$.

% \section{Executable Logical Scenarios Specification for Pedestrian Crossing}
% \label{sec:executable-semantics}
% \input{logical-scenarios}

\section{Experimental Results}
\label{sec:experiments}
%!TEX root = tase23.tex
\newcommand\pedCross{\ensuremath{\mathsf{pedCross}}}
\newcommand\pedCrBnds{\ensuremath{\mathsf{pedCrBnds}}}
\newcommand\pedCrErr{\ensuremath{\mathsf{pedCrErr}}}
\newcommand\senerr{\ensuremath{\mathsf{senerr}}}
\newcommand\folGap{\ensuremath{\mathsf{folGap}}}
\newcommand\folRSS{\ensuremath{\mathsf{folRSS}}}
\newcommand\maxLeadDec{\ensuremath{\mathsf{maxLeadDec}}}

We carried out a collection of experiments using 
the symbolic soft agents framework ~\cite{nigam22wrla}
and symbolic rewriting modulo SMT as described in Section~\ref{subsection:soft-agents}. Section~\ref{subsec:dt-experiments} describes the experiments for automatically proving $\dt$-adequacy. Section~\ref{subsec:timed-resilience-experiments} describes the experiments for automatically proving timed recoverability.
We used a value of $\dt = 0.1s$ for all experiments.  If an answer has not been returned after one hour, an experiment is aborted. All experiments were carried out on a 2.2 GHz 6-Core Intel Core i7 machine with 16 GB memory. The code is available in the folder rta\_symbolic\_agents at \url{https://github.com/SRI-CSL/VCPublic}.

We considered the scenarios described as follows:

% \paragraph{Summary of scenarios considered.}
\begin{itemize}
  \item \pedCross(\gapsafer,\gapsafe,\gapunsafe,\senerr) -- Pedestrian Crossing using only Relative Distances: This scenario is the pedestrian crossing scenario described in Section~\ref{sec:example}. The safety properties of the scenario are those as described in Equation~\ref{eq:ped_only_relative} using only relative distances and parametrized by the values \gapsafer, \gapsafe, \gapunsafe. We assume that the sensor that detects pedestrians and their properties, namely, speed, position and direction, may not be perfect. That is, the vehicle's local knowledge base, used to decide which action it will take, may not correspond to the ground truth.
  In particular, the position of the pedestrian inferred by the vehicle may differ by some 
amount proportional to the actual distance to
the pedestrian. 

\newcommand\err{\mathsf{err}}

The error, $\err$, is proportional to the distance ($\pos_p - \pos$) between the vehicle and the pedestrian as specified by the formula
$$
  \err \leq (\pos_p - \pos) \times \senerr \textrm { and } \err \geq 0. 
$$

\noindent
% \begin{verbatim}
%   (vv(i,"errX") <= (px - px0) * errPedS) and (vv(i,"errX") >= 0) 
%   px <= px0 + vv(i,"errX") and px >= px0 - vv(i,"errX")
% \end{verbatim}
% \red{Explain formula.}
In this case the safe controller of the vehicle is conservative, \eg, reducing the speed of the vehicle more aggressively, so to still satisfy the timed recoverability property.
When $\senerr = 0$, then the sensors are not faulty.
\item \pedCrBnds(\gapsafer,\gapsafe,\gapunsafe,\senerr) --  Pedestrian Crossing with \safer\ specified using low speeds and great distances: This is similar to the previous case, but now we are using the safety property for \safer\ as specified by Equation~\ref{eq:sp_ped_bounds}.

\item \folRSS($\maxdec_l$) -- Vehicle Following with RSS Properties: This scenario involves the vehicle following scenario using the safety properties based on the RSS property~\cite{rss} described in Example~\ref{ex:rss}.
We parametrize the safety property according to the assumed maximum deceleration of the leader ($\maxdec_l$). 
We follow the analysis carried out in~\cite{ensemble}. 
This work identifies three scenarios based on the expected occurrence of leader vehicle deceleration. 
The first scenario, which is highly unlikely, is that the leader makes an emergency brake ($\maxdec_l = -8 m/s^2$); 
the second when the leader vehicle decelerates heavily ($\maxdec_l = -5 m/s^2$); and the most likely case when the leader vehicle decelerates normally ($\maxdec_l = -2 m/s^2$).

\item \folGap(\gapsafer,\gapsafe,\gapunsafe) -- Vehicle Following with Gap Distances Properties: This scenario is described in more detail in~\cite{nigam22wrla}. In particular, we use safety properties based gap distances, similar to the pedestrian crossing.
% and described in~\cite{nigam22wrla}.
\end{itemize}

% The experiments  illustrate how the safety properties impact the performance of automating checking for these properties. Often the key limitation is the SMT-solver (Z3) that is not capable of checking within the experiment time of one hour the unsatisfiability of formulas. 

\subsection{Automating Recoverability Proofs using Symbolic Soft-Agents}
\label{subsection:soft-agents}

\begin{figure}[t]
\begin{center}
 \includegraphics[width=0.98\textwidth]{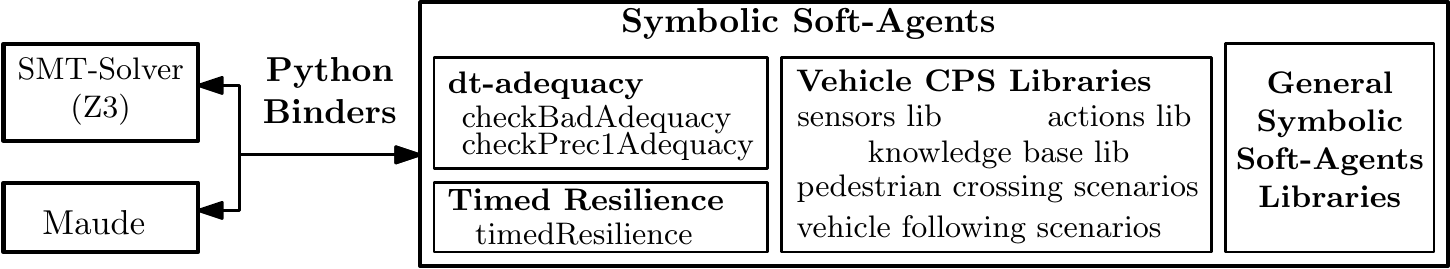}
\end{center}
\vspace{-5mm}
\caption{Key libraries and tools used for automating recoverability proofs.} 
\label{fig:symbolic-sa}
\vspace{-4mm}
\end{figure}

Figure~\ref{fig:symbolic-sa} depicts the main machinery that has been implemented and used. 
It is based on the soft-agents framework~\cite{talcott16quanticol}
and the general symbolic libraries described in~\cite{nigam22wrla}. The general symbolic soft-agents libraries specify the executable semantics of CPS based on rewriting rules.  The symbolic soft-agents rewrite rules correspond directly to the two \LS relations $\to_\tasks$ and $\to_\dt$.
We implemented the vehicle-specific libraries for specifying vehicle scenarios. 
% The Appendix contains the specification necessary for the pedestrian crossing scenarios, while we extend the specification in~\cite{nigam22wrla} for the vehicle following scenarios.
We have also implemented the machinery for checking for $\dt$-adequacy (\bad-adequacy and $\prec_1$-adequacy) and Timed Recoverability. 

The symbolic soft-agents are executable specifications. In particular, the execution traces are enumerated by Maude~\cite{clavel-etal-07maudebook} search.
The constraints in the traces (non-linear arithmetic formulas) are solved by the SMT-solver (Z3~\cite{z3}).
We  implemented the connection between the symbolic soft-agents libraries and SMT solvers using the Python Binders described in \cite{rubio22wrla}, thus enabling easy extensions to additional solvers and other tools in the future.

The basic idea
is to search for a counter-example to the property of interest.  Because the symbolic search is complete, failure to find
a counter-example means that the property holds for all
instances of the $\LS$ under consideration. \footnote{
In our scenarios, we consider traces of finite length. There may be infinitely may state, but search terminates as the states finitely represented using
symbolic terms.}

%%%%%%%%%CLT ADDED

\newcommand{\asys}{{\ensuremath{\mathsf{asys}}}\xspace}
\newcommand{\enforce}{{\ensuremath{\mathsf{enforce}}}\xspace}
\newcommand{\cond}{{\ensuremath{\mathsf{cond}}}\xspace}

As an example, to check bad-adequacy, the algorithm follows Proposition~\ref{prop:prec1} by searching for a
counter example, \ie\ properties $\SP_0 \prec_1 \SP_1$ (not \bad) and
$\LS$ instances $\conf_0$, $\conf_1$ such that $\conf_0$
satisfies $\SP_0$, $\conf_1$ satisfies $\SP_1$, 
$\conf_0 \to_\dt \conf_1$, 
and there is $\dt_0$ with $0 < \dt_0 < \dt$, $\conf_2$ such that $\conf_0 \to_{\dt_0}  \conf_2 \to_{\dt-\dt_0} \conf_1$,
where $\conf_2$ satisfies \bad.
If no counterexample is found then bad-adequacy holds for the
given \LS, \dt, and property specification.

Using symbolic rewriting, an arbitrary instance of \LS is
represented by a term, \asys, consisting of a
symbolic agent configuration and a symbolic environment. The
environment contains knowledge of the physical state and the
constraint on symbol values. The assertion that a property \SP holds
for a configuration is represented by the term 
$\enforce(\asys,\SP)$
that conjoins the boolean term specifying \SP in terms of the
symbols of \asys to the constraint in the environment. 
$\cond(\asys)$ is the constraint in the environment part of \asys.

The base case is adequacy for a pair of properties, 
$\SP_0$, $\SP_1$.
The algorithm for this case does the following.
First, use symbolic search from $\asys_0 = \enforce(\asys,\SP_0)$ for some $\asys_1$ such that  $\asys_0 \to_\dt \asys_1$ and 
  $\cond(\enforce(\asys_1,\SP_1)))$ is satisfiable. 
If no such $\asys_1$ is found, dt-adequacy holds for the
given $\SP_0,\SP_1$.
Otherwise, for some found $\asys_1$ do a symbolic search from (a copy of) 
$\asys_0$ for
some  $\asys_2,\dt_0$, where $\dt_0$ is  symbolic, 
such that $\asys_0 \to_{\dt_0} \asys_2$ and

\(
 \cond(\enforce(\asys_1,\SP_1)) \wedge \cond(\enforce(\asys_2,\bad))
   \wedge  0 < \dt_0 < \dt
\)

\noindent
is satisfiable.
If such $\asys_1, \asys_2,\dt_0$ are found we have a counter-example,
otherwise bad-adequacy holds for $\SP_0,\SP_1$.

The remaining algorithms for $\prec_1$-adequacy, noSkip property, and $t$-recoverability follow the same pattern as for bad-adequacy.

% This takes care of the 0 or 1 level property jumps.  
% The noSkip property must be checked to cover the remaining cases.

\subsection{\dt-adequacy Experiments}
\label{subsec:dt-experiments}

\begin{table}[t]
\begin{center}
  \begin{tabular}{p{3.3cm}p{2.5cm}p{5.5cm}}
  \toprule
  \multicolumn{3}{c}{\textbf{Pedestrian Crossing Scenarios}}\\
  \toprule
  \textbf{Scenario} & \textbf{$\bad$-adequacy} & \textbf{$\prec_1$-adequacy} \\
  \midrule
  $\pedCross(3,2,1,0)$ & Yes (130s) & DNF\\
  $\pedCrBnds(3,2,1,0)$ & Yes (172s) & No (358s), failed case from \safe\ to \safer.\\
  $\pedCross(5,2,1,0)$ & Yes (89s)  & Yes(149s) \\
  $\pedCrBnds(5,2,1,0)$ & Yes (78s) & No(172s), failed case from \safe\ to \safer. \\
  % \toprule
  % \multicolumn{3}{c}{\textbf{Vehicle Following Scenarios}}\\
  % \toprule
  % \textbf{Scenario} & \textbf{$\bad$-adequacy} & \textbf{$\prec_1$-adequacy} \\
  % \midrule
  $\folGap(3,2,1)$ & DNF & Yes (1413s) \\
  $\folGap(6,4,2)$ & Yes (51s) & No (52s), failed case from \safe\ to \safe.\\
  $\folGap(7,5,1)$ & Yes (55s) & Yes (83s)\\
  $\folRSS(-8)$ & DNF & DNF \\
  $\folRSS(-5)$ & DNF & DNF \\
  $\folRSS(-2)$ & Yes (304s) & Yes (533s)\\
\bottomrule
  \end{tabular}
\end{center}
\caption{
Automated proofs for $\bad$ and $\prec_1$-adequacy for different scenarios. DNF denotes that the experiment was aborted after one hour. }
\label{tb:experiments1}
\vspace{-8mm}
\end{table}

Table~\ref{tb:experiments1} presents our main experiments for proving $\dt$-adequacy. 
Since for each scenario there are four levels of properties (\bad, \unsafe, \safe, \safer), there are ten cases to consider, \eg, the case from starting at a configuration satisfying $\safer$ and ending at another configuration satisfying $\safe$ and so on. 

\paragraph{Pedestrian Crossing Scenarios:}
The soft-agents machinery is able to prove \bad-adequacy in less than 3 minutes. However, for $\prec_1$-adequacy, 
the soft-agents machinery fails to return a result for the scenario $\pedCross(3,2,1,0)$ (without the explicit bounds). In particular, the SMT-solver cannot prove or find a counter-example within one hour. If we increase the values of \gapsafer and \gapsafe to $5$ and $2$, then the soft-agent machinery terminates positively. While it is hard to formally justify this as the SMT-solver applies several heuristics, this is, intuitively, expected as these new values result in more coarse safety properties.

Moreover, the $pedCrBnds$ scenarios do not satisfy the $\prec_1$-adequacy. 
In particular, it fails  one case, namely, from $\safe$ to $\safer$. 
This seems to suggest that one can merge $\safe$ and $\safer$ in the analysis of recoverability, as we are still able to detect transitions to the lower properties (\unsafe\ and \bad).

\paragraph{Vehicle Following Scenarios:}
Both sets of scenarios were challenging for the soft-agents machinery. Differently from the pedestrian crossing example, \folGap\ was easier to prove $\prec_1$-adequacy and not terminating for \bad-adequacy. 
Interestingly, when increasing the \gapsafer, \gapsafe, \gapunsafe bounds to 6,4, and 2, respectively, $\prec_1$-adequacy failed in the case from $\safe$ to $\safe$, but increasing further the values to 7,5 and 1, the proof is established.
This indicates that the value of 2 for $\gapunsafe$ is not adequate as the system is capable of traversing a configuration satisfying $\bad$ within a $\dt$.
For \folRSS, the soft-agents machinery was only able to prove both adequacy properties when assuming a maximum deceleration for the leader vehicle of $-2 m/s^2$. 
% It is intuitively expected that this case is easier than the other cases with $-5 m/s^2$ and $-8 m/s^2$ as the latter cases subsume the case when  the range of values admitted for the leader vehicle.

In summary, all the scenarios, except $\folRSS(-5)$ and $\folRSS(-8)$, the soft-agents machinery is capable of demonstrating automatically \bad\ and $\prec_1$ adequacy. 
The cases of $\folRSS(-5)$ and $\folRSS(-8)$ are more challenging and the investigation on how to improve the machinery or CPS modeling to handle them is left to future work. 

\subsection{Time-bounded Recoverability Experiments}
\label{subsec:timed-resilience-experiments}

\begin{table}[t]
\begin{center}
  \begin{tabular}{p{3.5cm}p{4cm}p{4cm}}
  \toprule
  \multicolumn{3}{c}{\textbf{$\tup{\safe,\safer,t}$-One-Recovery-Period}}\\
  \toprule
  \multicolumn{3}{c}{\textbf{Pedestrian Crossing Scenarios}}\\
  \toprule
  \textbf{Scenario} & \textbf{$t = 4$} & \textbf{$t = 5$}  \\
  \midrule
  $\pedCross(3,2,1,0)$ & No (34s) & No (115s)  \\
  $\pedCrBnds(3,2,1,0)$ & No (27s) & Yes (621s) \\
  $\pedCross(5,2,1,0)$ & No (27s) & No (93s) \\
  $\pedCrBnds(3,2,1,0.50)$ & -- & No (103s) \\
  $\pedCrBnds(3,2,1,0.33)$ & -- & No (104s) \\
  $\pedCrBnds(3,2,1,0.125)$ & -- & Yes (637s) \\
  $\pedCrBnds(3,2,1,0.1)$ & -- & Yes (734s) \\
  \toprule
  \multicolumn{3}{c}{\textbf{Vehicle Following Scenarios}}\\
  \toprule
  \textbf{Scenario} & \textbf{Recoverability}  \\
  \midrule
  $\folGap(3,2,1)$ & $t = 5$ & No (12s) \\
  $\folGap(6,4,2)$ & $t = 5$ & No (11s)\\
  $\folGap(7,5,1)$ & $t = 5$ & No (12s)\\
  $\folRSS(-5)$ & $t = 2$ & No (5s) \\
  $\folRSS(-5)$ & $t = 3$ & No (81s) \\
  $\folRSS(-5)$ & $t = 4$ & No (1126s) \\
  $\folRSS(-5)$ & $t = 2$ & Yes (38s) $\star$ \\
  $\folRSS(-2)$ & $t = 2$ & Yes (43s)\\
\bottomrule
  \end{tabular}
\end{center}
\caption{
Automated proofs for  Timed-Recoverability. The symbol $\star$ denotes that the experiment used a very aggressive controller. 
 As the scenario $\pedCrBnds(3,2,1,0)$ is not recoverable for $t = 4$, it is not necessary to carry out experiments for the scenarios marked with --.}
\label{tb:experiments2}
\vspace{-5mm}
\end{table}

Table~\ref{tb:experiments2} summarizes our main experiments for recoverability involving the pedestrian crossing and vehicle following scenarios.
Recall that the objective of $\tup{\safe,\safer,t}$-Recoverability is to prove that the safety controller is capable of reducing vehicle risk to $\safer$. 
\omitthis{
Therefore, we specified reasonable controllers that reduce the agent's speed whenever in risky situation.
We do not enter into the details of these specifications and refer to~\cite{nigam22wrla} for some example of specifications. }
For the experiments we generally used simple, rather cautious
controllers.  For the purpose of illustration, we specified two controllers for the vehicle follower scenarios: a non-aggressive safety controller and an aggressive controller. 
The latter always activates the emergency brake, \ie, maximum deceleration.
Finally, for each scenario, our machinery showed that $\dt$ does not skip $\safe$ (see Definition~\ref{def:skip}) in around one second.

\paragraph{Pedestrian Crossing Scenarios:}
The first observation is that one is not able to establish recoverability with the safety properties used for \pedCross. 
Our machinery returns a counter-example where the vehicle has very low speeds and is very close to the pedestrian crossing with distance around $0.5m$. This illustrates the importance of including the bounds to safety properties as done in \pedCrBnds as in Equation~\ref{eq:sp_ped_bounds}. 

For the scenario $\pedCrBnds(3,2,1,0)$, the safety controller always returns to a \safer\ risk situation after 5 ticks, but not 4 ticks. 
Notice that for $\pedCrBnds(5,2,1,0)$ this is no longer the case as it fails also after 5 ticks. This is expected as the ``distance'' between the properties \safe\ and \safer\ has increased.

Finally, the experiments for \pedCrBnds(\gapsafer,\gapsafe,\gapunsafe,\senerr) illustrate how to check the recoverability of safety controllers in the presence of faulty sensors. 
If we assume faults of $50\%$ or $33\%$ on the pedestrian sensor, the safety controller cannot guarantee that it will always return to a \safer\ risk condition. However, it is able to do so for errors of $12.5\%$ or $10\%$.

\paragraph{Vehicle Following Scenarios:}
Our experiments demonstrate that it seems harder to establish recoverability when using time gaps to establish levels of risk. 
It probably requires a more sophisticated safety controller. 
On the other hand, when using RSS-based properties, it is possible to establish recoverability, even with small time frames, albeit when assuming normal decelerations of the leader vehicle.
It is possible to establish recoverability for scenarios assuming higher values for deceleration, but then a more aggressive controller is required.

\section{Related Work}
\label{sec:related}

% \begin{itemize}
% %	\item SOTER
% %	\item Mathias Althoff, Simulation-based (Verifai)
% 	% \item Check out the SOTIF paper [Vivek]
% %	\item Keymaera [Andre Platzer]
% \end{itemize}

%!TEX root = tase23.tex

\paragraph{\RAPs .} Since the first proposal of \RAPs, called Simplex Architecture~\cite{simplex2}, there has been several recent proposals of \RAP variants~\cite{ramakrishna20jsa,mehmood22nfm,damare22rv} (to name a few). 
While there are some differences on their architectures and functions, they all contain a decision module that evaluates the system risk level to decide which controller to use (the safe or the advanced controller). 
Therefore, all the requirements formalized in this paper, namely, the time sampling adequacy and recoverability are still relevant and applicable.
Indeed, we advance the state of the art by providing suitable definitions that are amenable to automated verification.

% \cite{simplex} specifies a way to synthesize a correct Simplex, but only when the system dynamics, thus not suitable for complex autonomous CPSes. More recent work has addressed $\RAP$ correctness for such systems~\cite{shankar,mehmood22nfm}.  

We have been inspired by~\cite{shankar} that proposes high-level requirements on the recoverability of \RAPs based on the level of risk of the system.
In particular, the methods for checking adequacy
of the sampling and for checking t-recoverability correspond
to the safety and liveness requirements of RTA wellformedness.  The third condition concerns the minimum
time to become unsafe (non-safe) with any controller in charge, needed to ensure that the monitor can switch controllers and the safe controller can react before reaching
an unsafe condition.  This can be shown using $<_1$-adequacy
and continuity of properties in a $<_1$-chain.
Summarizing, symbolic rewriting combined with SMT solving provides automated methods to verify correctness of time sampling mechanisms and safety requirements such as those of the RTA framework of \cite{shankar}.

In a similar direction, \cite{mehmood22nfm} proposes high-level requirements for the correctness of the decision module based on the definition of what is safe and existence of ``permanently safe command sequences'', which seems related to our time recoverability property.
They do not investigate, however, the effect of the time sampling and the correctness of the decision module.

\paragraph{CPS Verification and Validation}
Much of the literature in CPS verification, \eg~\cite{fremont20itsc} to name one, including some of the previous work on \RAP~~\cite{ramakrishna20jsa,damare22rv,mehmood22nfm}, rely on simulation-based methods.
These approaches are complementary to the one introduced in this paper. While this paper's approach targets more early phase development by providing proofs that \RAP specifications are suitable for all instances of a logical scenario, simulation-based approaches focus on later approaches for validating and testing implementations of \RAP systems on particular instances of logical scenarios.

\paragraph{dL, KeYmaera X, and VeriPhy.}
The KeYmaera X prover 
\cite{keymaerax,quesel-platzer-16tutorial} uses
differential dynamic logic (dL) to specify and
verify CPS controller designs. 
It is the starting
point of the VeriPhy pipeline 
\cite{bohrer-etal-18pldi-veriphy,bohrer-platzer-19waypoint} 
for producing code from logical specifications.
dL specifications and logical scenarios have in
common that they are given by terms with
constrained variables representing all instances
where values of variables satisfy the given
constraints. 
Our methods differ in that dL specifications are not directly
executable and therefore, one uses interactive theorem proving
methods to verify dl specifications, whereas logical scenarios
are executable thus enabling further automation of
verification proofs using rewriting modulo SMT.

\omitthis{
dL is a real-valued first-order dynamic logic
where the actions of modalities are hybrid
programs (HPs). a program notation for hybrid
automata. HPs mix discrete updates with system
evolution according to differential equations,
using constraints to determine when a
non-deterministic choice is allowed.  
The HPs
analyzed have the form (ctl ; plant)* where ctl is
the cyber part, setting control variables, and
plant is the physical part that evolves the
physical state variables according to a
differential equation specification for some
amount of time. This is similar to Soft Agent where
there are two rules, one that executes the agent
decision process and sets actions and one that
applies the actions to evolve the state over a
specified amount of time.

dL specifications and logical scenarios have in
common that they are given by terms with
constrained variables representing all instances
where values of variables satisfy the given
constraints. They differ in that dL specifications
are not executable while logical scenarios
are executable by symbolic rewriting.
}

\omitthis{dL
specifications can assert safety or liveness
properties and are verified by transformation
using KeYmaera proof rules augmented with constraint solving. Safety and resilience
properties of logical scenarios are verified by
reachability analysis using rewriting modulo SMT.
}

\paragraph{Formal Definitions of Resilience:}
Alturki \etal~\cite{alturki22ictac} propose formal definitions for resilience and show them to be undecidable in general and PSPACE-complete for some cases. 
While formal connections are left to future work, our definition of timed recoverability seem to specialize their definition so to be applicable for \RAP architectures, \eg, considering $\dt$-adequacy.

\vspace{-2mm}
\section{Conclusions}
\label{sec:conc}
%!TEX root = tase23.tex

In this paper we present methods to automate proving safety properties using abstract logical scenarios ($\LS$). An $\LS$
consists of instances of a pattern satisfying given ODD
constraints, together with a two-step transition relation giving
the semantics. The first step corresponds to reading sensors,
analyzing and deciding on actions (setting control parameters). The
second step evolves the system for the sampling time between
observations. Towards a formal foundation we introduce a notion of
Safety Property Specification for an $\LS$ as a set of property
(names) with a risk level ordering relation, a unique least (most
risky) element, $\bad$, and a satisfaction relation.  An adequate
sampling time should ensure that nothing important is missed.
We define three notions of $\dt$ adequacy and show that they
are distinct and totally ordered.  A system may be allowed to
enter a situation that is safe but risky, but a resilient system
will recover to an acceptably safe situation.  This is formalized
in a definition of $t$-recoverability.   A notion of 
one-period-recovery $t$-recoverability is defined that is amenable to verification, and
shown to be equivalent to $t$-recoverability for adequate $\dt$
using an inductive argument.

Towards automation of proofs, we use symbolic rewriting modulo SMT
as the execution and search engine \cite{nigam22wrla}.  Algorithms
were developed to prove all (infinitely many) instances of an $\LS$ satisfy different notions of $\dt$ adequacy or $t$-recoverability 
(or to provide counter example instances).  We report a set of
experiments checking $\dt$ adequacy and $t$-recoverability properties
for $\LS$s and safety property specifications related to vehicle automation: vehicle following and pedestrian crossing.  The experiments
show that it possible to find values of $\dt$ and safety parameters
where adequacy holds and very simple controllers satisfy $t$-recoverability.
They also highlight corner cases where things go awry.

One direction of future work is to investigate a wider range of
case studies to better understand how the different design
parameters interact.  Another important direction is to develop
methods to compose Logical Scenarios and proofs, thus scaling
analysis of complex systems.

\paragraph{Acknowledgments.} 
% % We thank Yuri Gil Dantas for valuable comments on a earlier version of this paper.
Talcott was partially supported by the U. S. Office of Naval Research under award numbers N00014-15-1-2202 and N00014-20-1-2644, and NRL grant N0017317-1-G002. We also thank anonymous reviewers for their valuable comments in earlier versions of this document.

 \newpage
 \bibliographystyle{abbrv}
 \bibliography{master}

\appendix
%!TEX root = tacas22.tex
\section{Proof of Proposition \ref{prop:prec1}}
Assume the conditions of the proposition with $\conf$ an instance of $\LS$
and $\conf \to_{\dt'} \conf' \to_{\dt-dt'} \conf_1$.  There are two cases.
First, assume that $\SP_1 \neq \SP_2$. 
From the $\prec_1$ property, we have $\conf' \vDash \SP_1$ or 
$\conf' \vDash \SP_2$. Since $\SP_1 \neq \bad$ and $\SP_2 \neq \bad$, 
and properties are disjoint, $\conf' \nvDash \bad$.
Second, assume that $\SP_1 = \SP_2$.
From the $\prec_1$ property, we have $\conf' \vDash \SP_1$ or 
$\conf' \vDash \SP$ for some property $\SP$ such that  
$\SP \prec_1 \SP_1$ or $\SP_1 \prec_1 \SP$. Since $\bad \neq \SP_1$ 
and $\bad \nprec_1 \SP_1$,  $\conf' \nvDash \bad$.

\section{Proof of Proposition \ref{prop:bad-dt-adeq}}
  Assume $\dt$ is one transition adequate. 
  Let $\conf \rightarrow_\dt \conf_1$ be an arbitrary transition.
  There are two cases to consider for $\prec_1$-adequacy:

  Case 1: $\conf \vDash \SP_1$ and $\conf \vDash \SP_2$ with $\SP_1 \neq \SP_2$, so that for any decomposition $\conf \to_{\dt'} \conf_2 \to_{\dt-\dt'} \conf_1$, we have $\conf_2 \vDash \SP_1$ or $\conf_2 \vDash \SP_2$. 
  From $\dt$ one transition adequacy, we know that there is a $0 < \dt'' < \dt$, that is the transition from $\SP_1$ to $\SP_2$. 
  We can split into two sub-case:
  \begin{itemize}
    \item Case 1.1: Let $0 < \dt' < \dt''$. From $\dt$ one adequacy,
     all $\conf_2$ where $\conf \to_{\dt_2} \conf_2 \to_{\dt''-\dt_2}
      \conf_d$, 
      we have that $\conf_2 \vDash \SP_1$. Therefore, all reachable 
      configurations $\conf'$ with $\dt'$, we have $\conf'\vDash \SP_1$;
    \item Case 1.2: Let $\dt'' \leq \dt' < \dt$. The reasoning is similar, but with $\SP_2$, resulting in $\conf'\vDash \SP_2$
  \end{itemize}
  The second case when $\conf \vDash \SP$ and $\conf \vDash \SP$ follows the same reasoning, but where $\SP_1 = \SP_2 = \SP$.

  For $\bad$-adequacy, assume that $\conf \vDash \SP_1$ and $\conf_1 \vDash \SP_2$ with $\SP_1 \neq \bad$ and $\SP_2 \neq \bad$. 
  From one transition adequacy, all the intermediate configurations, $\conf'$  either $\conf' \vDash \SP_1$ or $\conf' \vDash \SP_2$, which implies that $\conf' \nvDash \bad$.

  Finally, assume that $\dt$ is $\prec_1$-adequate. 
  Let $\conf \vDash \SP_1$ and $\conf \vDash \SP_2$. 
  Let  $\conf \to_{\dt'} \conf' \to_{\dt-\dt'} \conf_1$ be an arbitrary decomposition.
  Assume that for $i \in \{1,2\}$ $\bad \nprec_1 \SP_i$ and $\bad \neq \SP_i$.
  Then there are two subcases. Case 1) If $\SP_1 \neq \SP_2$, from $\prec_1$ adequacy, we have that all $\conf' \vDash \SP_1$ or $\conf'\vDash \SP_2$.
  Case 2) If $\SP_1 = \SP_2$, from $\prec_1$-adequacy, $\conf' \vDash \SP'$ with either $\SP'\prec_1 \SP_1$ or $\SP_1 \prec \SP'$. In both cases $\SP'\neq \bad$.

\section{Proof of Theorem \ref{th:resilience}}
Assume
system is one-recovery period resilient.  
Let $\conf_0 \to_\tasks \conf' \to_{\dt} \conf_1 \to_\tasks \conf_1' \to_{\dt} \to \conf_2 \cdots$ be an arbitrary trace such that $\conf_0 \vDash \SP$ with $\SP = \SP_\safer$ or $\SP_\safer \prec \SP$.
We can classify configurations in the trace $\conf_0 \to_\tasks \conf' \to_{\dt} \conf_1 \to_\tasks \conf_1' \to_{\dt} \to \conf_2 \cdots$ as part of a normal operation mode and part of a recovery period mode.
In particular, we start by classifying states as normal operation mode, then whenever $\conf_k \vDash \SP_\safe$, we classify the configuration as recovery mode until one reaches a $\conf_l \vDash \SP_\safer$ with $l > k$. Such $\conf_l$ is guaranteed to exist due to the resilience property. At this point we classify again configurations as normal operation.
The segment from $\conf_k$ to $\conf_l$ defines a recovery period. The remaining segments are normal operation periods.

Now, the proof goes by induction on the number of recovery and normal periods. 
All configurations in a normal period satisfy a property $\SP$ with $\SP_\safe \prec \SP$, thus not satisfying $\bad$. Moreover, from $\dt$ adequacy, no decomposition satisfies $\bad$ either. 
On a recovery period, no decomposition satisfies $\bad$ from the definition of one-period-recovery resilience. (In fact no decompositions satisfy \bad.)
Moreover, from the fact that $\dt$ does not skip $\SP_\safe$, it is not possible not to activate a recovery period.

For the other direction, assume that the system is resilient. 
This means in particular that all traces starting with a $\conf$ such that $\conf \vDash \SP_\safe$ will recover within $t$ time units and no decomposition will pass through a configuration satisfying $\bad$. That is, the first piece of the trace is already a one-period recovery. 
This means that the system is also one-period-recovery resilient. 

\section{Specification of Pedestrian Crossing in Soft-Agents}
%!TEX root = tacas22.tex
Following the methodology described in~\cite{nigam22wrla} for automatically generating  safety proofs for CPSes, we have implemented the machinery necessary for specifying logical scenarios such as the pedestrian crossing described above.
\omitthis{
We use this example to describe the methodology and refer to our previous work~\cite{nigam22wrla} for the formal, precise details.}
Specifications are written in Maude~\cite{clavel-etal-07maudebook}, which implements rewriting logic~\cite{meseguer-92unified-tcs}. 
While familiarity with Maude is helpful, we believe that the specifications are understandable without it.
% The code is available at \red{XXX}.

\begin{figure}[t]
\begin{center}
  \includegraphics[width=0.75\textwidth]{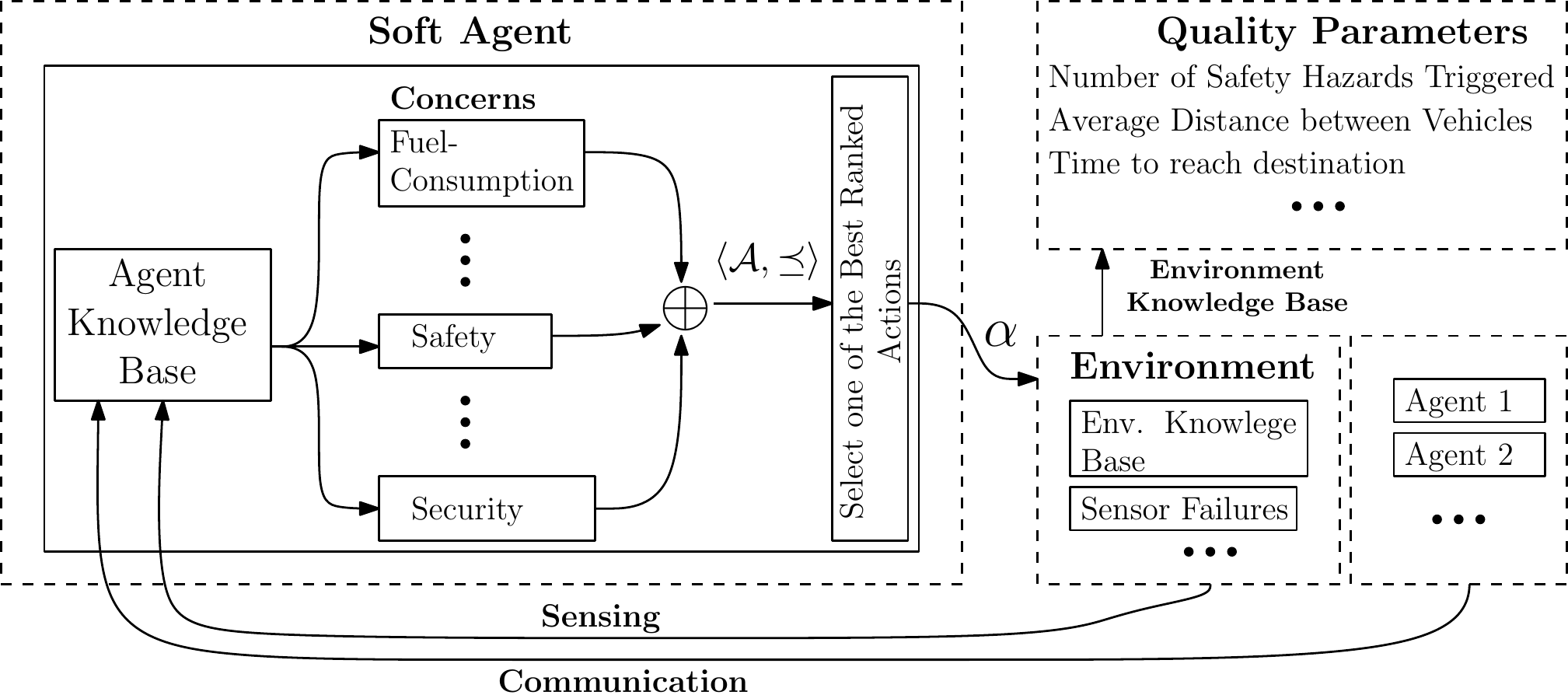}  
\end{center}
\vspace{-5mm}
  \caption{Soft Agent (\SA) architecture.}
  \label{fig:soft-agent}
\end{figure}

Figure~\ref{fig:soft-agent} depicts the overall architecture of a Soft Agent (\SA). It adopts a traditional sense-understand-decide-act architecture. In particular, a \SA has a local knowledge (\lkb) which specifies \SA's understanding of the environment. \lkb is  updated according to information provided by reading sensors. The sensors may provide wrong information, that is, information not corresponding to the environment. 
Based on \lkb, the \SA ranks its set of actions, \eg, accelerate or decelerate, according to different concerns. For the safety controller, safety is the main concern. Finally, the \SA carries out an action that has the maximum rank. 

The whole cycle corresponds to a logical time-tick. A logical time-tick corresponds to a real number. For an \RAP instance, the time-tick corresponds to the sample interval \dt, as the \RAP shall sense-understand-decide-act within \dt time units.

\subsection{Symbolic Knowledge Specifications}
\label{subsection:knowledge}
The main innovation of our previous work~\cite{nigam22wrla} is that instead of using concrete values for, \eg, location and speeds, as done previously~\cite{dantas20vnc}, we now represent these values by constrained symbols of sort $\RealSym$. 
Elements of $\RealSym$ are either real number constants, or symbols of the form \texttt{vv(i)} or \texttt{vv(i,str)} where \texttt{i} is a \texttt{Nat} uniquely identifying a symbol and \texttt{str} is a string describing the intuitive meaning of the symbol,  used for improved readability. 

The following symbols represent the initial conditions  for \veh, namely, its position, speed, maximum acceleration, maximum deceleration, and initial acceleration.
\begin{small}
\begin{verbatim}
 eq vposx = vv(2,"vh-posX") .      eq vposy = vv(3,"vh-posY") . 
 eq vvel = vv(5,"vh-speed") .      eq maxacc = vv(9,"vh-maxAcc") .
 eq maxdec = vv(10,"vh-maxDec") .  eq acc = vv(32,"vh-acc") .
\end{verbatim}   
\end{small}
The values that these symbols may represent are specified by a set of (non-linear) arithmetic constraints. 

\begin{example}
The following constraints specify the values for the vehicle's initial speed and acceleration as defined in the pedestrian crossing ODD:
\begin{small}
\begin{verbatim}
baseCond = {2 <= vvel and  vvel <= 10 
            and acc >= maxdec and acc <= maxacc, 
            maxacc == 2 and maxdec == -8}
\end{verbatim}   
\end{small}
These constraints specify, \eg, that the vehicle speed can be any value between 2m/s and 10m/s exactly as specified in the pedestrian crossing logical scenario. 

The initial knowledge base of the  vehicle, \veh, at logical tick 0, contains the following terms, specifying its initial position, speed, acceleration, and direction: 
\begin{small}
  \begin{verbatim}
  lkbI = (atloc(vh,loc(vposx,vposy)) @ 0) (speed(vh,vvel) @ 0) 
         (accel(vh,acc1) @ 0) (dir(vh,loc(0,0),loc(0,1),1) @ 0)
\end{verbatim}
\end{small}
The vehicle's direction is a vector specified by two points and a magnitude. Initially, the vehicle is moving on the Y-axis.
Moreover, \veh has not yet detected the pedestrian.
\label{ex:baseCond}  
\end{example}

An \SA updates its local knowledge base by observing the environment through its sensors. For example, \veh has the set of sensors \texttt{sset}:
\begin{verbatim}
eq sset = (locS speedS pedS(pedSDist,errPedS)) . 
\end{verbatim}
where \texttt{locS} is a location sensor, \ie, determines \veh's position; \texttt{speedS} its speed; and \texttt{pedS} detects pedestrians. \texttt{pedS} has two parameters. The first, \texttt{pedSDist}, specifies the distance in which a pedestrian can be detected.  The second parameter, \texttt{errPedS}, specifies the maximum relative error on the position of the pedestrian. These parameters abstracts issues such as camera resolution, that may lead to error or non detection of objects.

During execution at logical time \texttt{t}, sensors are used to update the information in \SA's local knowledge base. For example, if the pedestrian \ped is within sensing range, the \texttt{pedS} includes a knowledge item of the form: 
\begin{verbatim}
  (ped(p1,loc(px,py),spd1,loc(px1,py1),loc(px2,py2) @ t)
\end{verbatim}
where \texttt{loc(px,py)} is the sensed \ped location, which may be erroneous; \texttt{spd1} its speed; and the pair \texttt{loc(px1,py1),loc(px2,py2)} specifies its direction. 
Since the pedestrian sensor may be erroneous, as specified by the parameter \texttt{errorPedS}, the sensed pedestrian location \texttt{loc(px,py)} may not correspond to the pedestrian actual position. 
This is also specified symbolically by means of constraints.

\begin{example}
\label{ex:errsensor}
The constraint below, called \ErrSensor, specifies the sensed pedestrian X-axis position, where \texttt{loc(vx,vy)} and \texttt{loc(px0,py0)} are, respectively, the actual positions of the vehicle and of the pedestrian:
\begin{verbatim}
 (vv(i,"errY") <= (vy - py0) * errPedS) and 
 (vv(i,"errY") >= 0) 
 py <= py0 + vv(i,"errY") and py >= py0 - vv(i,"errY")
\end{verbatim}
The symbol \texttt{vv(i,"errY")} is a fresh symbol specifying the maximum error on the Y-axis. A similar constraint can be used for specifying an error for the X-axis. However, since the logical scenario, the vehicle is moving on the Y-axis, this is not needed.   
\end{example}

\subsection{Agent, Environment, System Configurations and Executable Semantics for Pedestrian Crossing}
\label{subsection:executablesemantics}
An agent configuration has the form \texttt{[id : class | attrs ]}, where \texttt{id} is the agent's unique identifier, \texttt{class} is its class, \eg, vehicle (\texttt{veh}), and \texttt{attrs} are its attributes which include its local knowledge base written \texttt{lkb : kb}, where \texttt{lkb} is a label and \texttt{kb} is the local knowledge base contents, and include sensors \texttt{sensors : sset}, where \texttt{sset} is a collection of sensors including \texttt{pedS(pedSDist,errPedS)} for sensing pedestrians.

An environment configuration has the form \texttt{[eId | ekb]} where \texttt{ekb} is the environment knowledge base which specifies state of the world. 
The environment knowledge base contains the knowledge item \texttt{constraints(i,cond)} where \texttt{i} is the current index of fresh variables, and \texttt{cond} is the constraints (accumulated) on the existing symbols.  

A system configuration is then a collection of agent configurations and an environment configuration. For example, a configuration specifying the pedestrian crossing logical scenario is as follows:
\begin{small}
 \begin{verbatim}
  asysI = { [eid | (kb constraint(i,condI))]
        [vh : veh | lkb : kb0 ; sensors : sset] 
        [p1 : ped | lkb : kb1 ] }
\end{verbatim} 
\end{small}%
It contains the agent configuration for the vehicle \veh and for the pedestrian \ped. 

The set of (initial) constraints \texttt{condI} specify the instances of a logical scenarios. 
The following are some examples of how constraints can be used to specify assumptions about the logical scenario.

\begin{example}
 The following set of initial constraints \texttt{condI} specifies a logical scenario with exactly one instance.
\begin{verbatim}
  vposx == 5 and vposy == 0 and vvel == 5 and acc == 1
  st == 30 and fn == 45 and pvel == 1 
\end{verbatim}
where the vehicle's initial position is (0,5), with a speed of $5m/s$, acceleration of $1m/s^2$, and the pedestrian is walking from the position 30 of the road edge to to the position 45 of the other side of the road with constant speed of $1m/s$.
\end{example}

More interesting is when the logical scenario denotes an uncountable number of instances.  
  This is the case with the following two examples which also assumes \texttt{baseCond} in Example~\ref{ex:baseCond}.
\begin{example}
Together with \texttt{baseCond}, the following constraints
\begin{verbatim}
vposy >= 0 and vposy < cr1 and 
           st == fn and st >= cr1 and sn <= cr2
\end{verbatim}
specify that the vehicle is facing the pedestrian crossing and from \texttt{st == fn}, the pedestrian is moving on the X-axis within the pedestrian crossing.

By removing the constraint \texttt{st == fn} and adding \texttt{fn >= cr1} and \texttt{fn <= cr2}, the resulting constraints specifies instances where the pedestrian is crossing (in straightline) from any point of the crossing at one side of the road to any point of the crossing on the other side of the road.
\end{example}

\begin{example}
Finally, we can also specify scenarios with different assumptions on the error of pedestrian detection sensors. For example, if one adds the constraint \texttt{errPedS == 0}, one assume a perfect pedestrian sensor. The constraint \texttt{errPedS <= 0.5}, on the other hand, specifies on the other hand a maximum error of 50\% on the relative distance to the pedestrian (as described above). 
\end{example}

\paragraph{Executable Semantics}
A key feature of the framework is that such symbolic system configurations are executable. 
The semantics is specified by two steps, formalized as rewriting rules over system configurations. 
\begin{itemize}
  \item In the first phase, agents update their local knowledge bases with the information obtained by the their sensors, and decide which actions to take, \eg, accelerate or decelerate.
  \item Then the second phase  executes the actions decided by the agents. 
  This means generating fresh symbols for the positions, speeds and other physical attributes of agents, and adding new constraints for these fresh symbols according to the expected physics.
\end{itemize}

For example, consider \veh's state as specified by the constrained symbols shown in Example~\ref{ex:baseCond}. Moreover, let \veh decide to maintain a constant speed, \ie, zero acceleration and not change its direction. Moreover, let \dt be the time corresponding to a logical tick.
The time tick rule will result in the following new knowledge base:
\begin{small}
  \begin{verbatim}
(atloc(vh,loc(vxnu,vynu)) @ 1) 
(speed(vh,velnu) @ 1) (accel(vh,accnu) @ 1) 
\end{verbatim}
\end{small}
and adds the followin constraints specifying \veh's possible state at logical time 1:
\begin{small}
  \begin{verbatim}
  vynu == vposy + (vvel + velnu) * dt / 2 and 
  velnu == vvel + accnu * dt and vxnu == vposx and accnu == 0
\end{verbatim}
\end{small}
Notice that the new configuration is also symbolic, \ie, correspond to a possibly uncountable number of concrete configurations.

% \red{TODO}
% \begin{itemize}
%   \item Decision Sampling-Time Completeness Problem
%   \item Time-Bounded Resilience Problem
% \end{itemize}

\end{document}